\documentclass[aps,prl,twocolumn,showpacs,10pt,superscriptaddress,a4paper,groupedaddress]{revtex4-1}
\usepackage{graphicx}
\usepackage{amssymb,amsmath}
\usepackage{epsfig}
\usepackage{txfonts}

\usepackage{amsmath,amssymb,amsfonts}
\usepackage{mathrsfs}
\usepackage{empheq}

\usepackage{color}
\usepackage[unicode,breaklinks]{hyperref}

\usepackage[unicode]{hyperref}
\hypersetup{
    pdftitle={Anomalous},
    pdfauthor={Yu, Bradley},
    pdfsubject={Vortex fluids},
    colorlinks=true,
    linkcolor=blue,
    citecolor=blue,
    filecolor=black,
    urlcolor=blue
}
\def\urlprefix{}
   \def\url#1{}

\newcommand{\EQ}[1]{\begin{align}#1\end{align}}
\newcommand{\eref}[1]{Eq.~\eqref{#1}}

\newcommand{\be}{\begin{equation}}
\newcommand{\ee}{\end{equation}}
\newcommand{\bea}[1]{\begin{align}#1\end{align}}

\newcommand{\nn}{\nonumber }

\newcommand{\lab}[1]{\label{#1}}

\newcommand{\sig}{\sigma}
\newcommand{\ome}{\omega}

\newcommand{\gam}{\gamma}

\newcommand{\rr}{\mathbf{r}}

\newcommand{\bz}{\bar z}
\newcommand{\del}{\partial}

\newcommand{\uu}{\mathbf{u}}
\newcommand{\vv}{\mathbf{v}}

\newcommand{\fref}[1]{Fig.~\ref{#1}}

\renewcommand{\Im}{\operatorname{Im}}

\begin{document}
\title{Emergent Non-Eulerian Hydrodynamics of Quantum Vortices in Two Dimensions}

\author{Xiaoquan Yu}
\email{xiaoquan.yu@otago.ac.nz}
\author{Ashton S. Bradley}
\email{ashton.bradley@otago.ac.nz}
\affiliation{Department of Physics, Centre for Quantum Science, and Dodd-Walls Centre for Photonic and Quantum Technologies, University of Otago, Dunedin, New Zealand. } 
\date{\today}

\begin{abstract}

We develop a coarse-grained description of the point-vortex model, finding that a large number of planar vortices and antivortices behave as an inviscid non-Eulerian fluid at large scales. The emergent binary vortex fluid is subject to anomalous stresses absent from Euler's equation, caused by the singular nature of quantum vortices. The binary vortex fluid is compressible, and has an asymmetric Cauchy stress tensor allowing orbital angular momentum exchange with the vorticity and vortex density. An analytic solution for vortex shear flow driven by anomalous stresses is in excellent agreement with numerical simulations of the point-vortex model.

\end{abstract}

\maketitle
\emph{Introduction.|}
Topological defects play a key role in the two-dimensional (2D) superfluid transition~\cite{Thouless72,Thouless73}, and in many out of equilibrium phenomena including phase ordering dynamics after a quench~\cite{Bray,Goldenfeld,Bray2000} and turbulence in Bose-Einstein condensates (BECs)~\cite{barenghi2001quantized,vinen2006introduction,barenghi2014introduction,Tsatsos:2016fu,Navon2016,anderson2016fluid,Tsubota2017,Henn09a}. Under planar confinement~\cite{footnote0} the collective motion of many interacting quantum vortices in a superfluid leads to the emergent complexity of 2D quantum turbulence (2DQT)~\cite{DEC2010,NeelyPRL,Matt2012,AshtonPRX,Chesler368,VortexDipolePRL,Relaxation2014Shin,Barenghi2015,IECMatt,TomPRA,AnghelutaPRE}. Experimental advances now allow plane-confined BECs~\cite{Dalibard2015,Gauthier16}, and simultaneous detection of quantum vortex positions and circulations~\cite{Shin2016}, opening the way to studies of 2DQT and Onsager vortices~\cite{Onsager1949,EyinkRMP,JM73,MJ74,DecayingQTBillam,TapioPRL,clusteringYu, Groszek:2015ty}.

In the hydrodynamic regime at low temperature, superfluid vortex motion  is much slower than the rate of sound propagation, and a system of many quantum vortices evolves as an almost isolated subsystem: a \emph{vortex fluid}. The point-vortex model (PVM), a central model for studying 2D classical incompressible turbulent flows~\cite{Onsager1949,EyinkRMP,novikov1975,Aref1999}, describes the dynamics of quantum vortices~\cite{Fetter67,HaldaneYongshi}, provided they are well-separated. In this regime the vortex core structure is unimportant, and the coupling to acoustic modes is relatively weak~\cite{Lucas}. Many studies of collective dynamics of vortices rely on large-scale numerical simulations of the discrete PVM~\cite{Numasato2009,siggia1981,Angheluta17,Matt2017}.
An alternate approach was initiated by Wiegmann and Abanov~\cite{Wiegmann}, who formulated a hydrodynamic description of well-separated vortices of the same circulation, providing a rigorous starting point for studying rotating fluids, vortex clusters with definite sign of vorticity~\cite{SmithPRL,SmithONeil}, and the connection between vortex fluids and quantum Hall liquids~\cite{VortexfluidHalleffect}. 
%Yet a theory of the collective motion for many planar vortices remains an open problem. 
A general 2D turbulent flow or a phase ordering process involves many vortices and anti-vortices, motivating a hydrodynamic theory of the binary vortex system.

In this Letter we develop a hydrodynamic formulation of systems involving a large number of vortices and anti-vortices, providing a framework for describing their emergent collective dynamics at large scales. A system containing many vortices with separation much bigger than the vortex core size $\xi$ can be treated as a fluid on a scale much larger than $\xi$, and smaller than the system size. We generalize the coarse-graining procedure proposed for chiral vortex systems~\cite{Wiegmann} to the binary vortex system by introducing two hydrodynamical velocity fields via vortex number and charge currents. The binary fluid obeys a compressible hydrodynamic equation containing an asymmetric Cauchy stress tensor and an anomalous stress that is analogous to viscous stress while conserving energy. A vortex fluid shear flow driven by the anomalous stress is found, and numerical simulations of the PVM show excellent agreement with the analytical solution.  Variation of the coarse-graining scale also demonstrates convergence of many-body dynamics of the PVM to non-Eulerian hydrodynamics.  Dissipation due to the interaction between vortices and thermal cloud generates uphill diffusion of the vorticity at macroscopic scales. Chiral flow dynamics~\cite{Wiegmann} is recovered as a special case of the binary fluid.

\emph{Two-Dimensional Hydrodynamics.|}
The dynamics of non-viscous incompressible classical fluids in 2D is described by the Euler equation  
\bea{\label{GPEuler}
	{\cal D}^{u}_t\mathbf{u}=-\frac{1}{n m}\nabla p, \quad \nabla \cdot \mathbf{u}=0,}
where $n$ is the constant density, $\mathbf{u}$ is the fluid velocity, ${\cal D}^{u}_t \equiv \del_t +\uu\cdot\nabla$ is the material derivative with respect to $\mathbf{u}$, $m$ is the atomic mass, and $p$ is the fluid pressure  determined by $ \nabla \cdot (\mathbf{u} \cdot \nabla \mathbf{u})=-(nm)^{-1}\nabla^2 p$. Taking the curl of \eref{GPEuler}, the Helmholtz equation for the vorticity $\omega\equiv [\nabla \times \mathbf{u}]_z$ is 
\bea{\label{Hw}{\cal D}^{u}_t\omega=0.} 
The kinetic energy of the fluid reads~\cite{footnote2}
\EQ{\lab{Hsf}
	H=\frac{nm}{2}\int d^2\mathbf{r}\;|\uu |^2=\frac{nm}{2}\int d^2\mathbf{r}\;\psi(\rr)\ome(\rr),}
where $\psi$ is the stream function, and ${\mathbf u}=\nabla \times (\psi \mathbf{\hat z})$ and 
$-\nabla^2 \psi=\omega$. Here $\mathbf{\hat{z}}$ is a unit normal vector to the fluid plane.

For a 2D flow, it is convenient to use complex coordinates $z=x+iy$, $\partial_z=(\partial_x-i\partial_y)/2$ and the complex velocity $u=u_x-iu_y$. 
In terms of complex notation 
$\nabla \cdot \mathbf{u}=\partial_z \bar{u}+\partial_{\bar{z}}u$, $\mathbf{u} \cdot \nabla=\bar{u}\partial_z + u \partial_{\bar{z}}$, $\omega=[\nabla \times \mathbf{u}]_z=i(\partial_{\bar z} u-\partial_z \bar u)=2i \partial_{\bar z} u$, and $u=2i \partial_z \psi$. We also use subscripts $a,b$ to denoted the Cartesian components of vectors, and use vector, complex, or component notation where convenient. 

\emph{Point-Vortex System.|} A superfluid containing vortices with separation larger than the core size $\xi$ is nearly incompressible, away from vortex cores~\cite{footnote11}. For a BEC described by a macroscopic wavefunction $\Psi$, the associated Gross-Pitaevskii equation (GPE) governing time evolution can be mapped to the form Eq.~\eqref{GPEuler} in the incompressible regime (constant density $n=|\Psi|^2$)~\cite{BECbook}.
The single valuedness of the wave function $\Psi$ requires that the circulation of a vortex excitation must be quantized in units of circulation quantum $\kappa \equiv2 \pi \hbar/m$, and the vorticity has a singularity at the position of the vortex core $\mathbf{r}_i$: $\omega(\mathbf{r})=\kappa \sigma_i \delta(\mathbf{r}-\mathbf{r}_i)$ with the sign $\sigma_i=\pm 1$. Hereafter we set $nm=1$ for convenience. 

We consider a system containing $N_+$ singly-charged quantum vortices and $N_-$ anti-vortices. The total number of vortices is  $N=N_+ + N_-$ and the vortex with sign $\sigma_i$ is located at $\mathbf{r}_i$. 
The fluid velocity generated by these quantum vortices far from the fluid boundary is completely determined by the vorticity: $u=2i \partial_z \psi=(2\pi i)^{-1} \int d^2\mathbf{r}'\; \omega(\mathbf{r}')/(z-z')=-\sum^N_{j=1} i \gamma \sigma_j/ (z-z_j)$,
where the stream function of the fluid is 
$\psi(\mathbf{r})=-\gamma \sum_i \sigma_i \log\left|(\mathbf{r}-\mathbf{r}_i)/\ell\right|$, and the vorticity is $\omega(\mathbf{r})=2 \pi \gamma \sum_i \sigma_i \delta(\mathbf{r}-\mathbf{r}_i)$.
Here $\ell$ is a length-scale introduced to ensure the correct dimension, and $\gamma=\kappa /2\pi$ is a convenient unit of circulation. As shown by Helmholtz, the above fluid velocity $u$ is a singular solution of Eq.~\eqref{Hw}. 
A quantum vortex generates a flow in the bulk fluid, while the vortex core is a point-like particle and has its own dynamics driven by the flow generated by the other vortices. 
The dynamics of the vortex cores has a Hamiltonian structure:
\EQ{\lab{Hamiltonianstructure}
	\frac{dz_i}{dt}=\frac{\del{\cal H}}{\del p_i},\hspace{.7cm}\frac{dp_i}{dt}=-\frac{\del{\cal H}}{\del z_i},
}
with canonical momentum 
$p_i=i\pi\gam\sig_i  \bz_i$,
and Hamiltonian 
\EQ{\lab{PointVortexH}
	{\cal H}=-\pi \gam^2\sum_{i\neq j}\sig_i\sig_j\log{\left|\frac{z_i-z_j}{\ell}\right|}.
}
The formal solutions of Eq.~\eqref{Hamiltonianstructure} are the Kirchhoff equations~\cite{Kirchoff} 
\bea{
	\label{Kirchhoff}\frac{d \bar{z}_i}{dt}&=v_i, \quad  v_i=-\sum^N_{j,j\neq i} \frac{i \gamma \sigma_j}{z_i(t)-z_j(t)}.
}
Using the vortex degrees of freedom, the kinetic energy of the fluid $H={\cal H}+E_{\rm self}$, where ${\cal H}$ is energy of interaction between vortices, and $E_{\rm self}=N\pi \gamma^2 \log (\ell/\xi)$ is the total self-energy.
The PVM can be seen as a limiting case of the vortex method~\cite{chorin,mathvortex} and well-approximates incompressible classical fluids with $\kappa $ determined by the injection scale.

\emph{Vortex Fluid Hydrodynamics.|}
For large $N$,  the emergent collective dynamics of the discrete vortex system Eq.~\eqref{Kirchhoff} can be described by a few hydrodynamic variables.
By coarse-graining microscopic vortex distributions over patches containing many vortices, we derive a hydrodynamic formulation of the PVM, describing the vortex dynamics on scales greater than the patch scale. The core size $\xi$ is much smaller than the patch scale, and serves as a natural ultraviolet cut-off.

Under Hamiltonian evolution, conservation laws ensure the following continuity equations
\bea{\label{continuity2}\partial_t \rho&=\sum_{i} \partial_t \delta (\mathbf{r}-\mathbf{r}_i(t))=-\left(\partial_z \bar{J}_{\rm n}+\partial_{\bar{z}}J_{\rm n}\right)=-\nabla \cdot \mathbf{J}_{\rm n}, \\
	\label{continuity1}\partial_t \sigma&=\sum_{i} \sigma_i \partial_t \delta (\mathbf{r}-\mathbf{r}_i(t))=-\left(\partial_z \bar{J}_{\rm c}+\partial_{\bar{z}} J_{\rm c}\right)=-\nabla \cdot \mathbf{J}_{\rm c},
} 
where vortex number density  $\rho(\mathbf{r})\equiv\sum_i\delta(\mathbf{r}-\mathbf{r}_i)$, vortex charge density $\sigma(\mathbf{r})\equiv\sum_i\sigma_i \delta(\mathbf{r}-\mathbf{r}_i)=(2\pi\gamma)^{-1}\omega$, and
\bea{\label{micJ}
	J_{\rm c}&=\sum_i \delta(\mathbf{r}-\mathbf{r}_i) (\sigma_i v_i), \quad   J_{\rm n}=\sum_i \delta(\mathbf{r}-\mathbf{r}_i) v_i,
}
are the currents for charge and number respectively.
The coarse-grained vortex charge velocity field $w$ and  vortex velocity field $v$ are defined according to the hydrodynamic relations
$J_{\rm c} \equiv \rho w $ and $J_{\rm n} \equiv \rho v$~\cite{footnote9}. 
Using the identity
\bea{\label{decompozationbinary} 2\sum_{i\neq j} \frac{\sigma_i}{z-z_i} \frac{\sigma_j}{z_i-z_j}
	=\left(\sum_{i} \frac{\sigma_i}{z-z_i}\right)^2+\partial_{z}\sum_{i}\frac{1}{z-z_i}}
with \eref{Kirchhoff}, we can rewrite $J_{\rm c}$, to obtain an important relation between the vortex charge velocity $w$, and the fluid velocity $u$, given by
\bea{\label{canonical}\rho w = \sigma u-2\eta i\partial_z \rho,}
with anomalous kinetic coefficient $\eta=\gamma/4$; here we have used $\partial_{\bar z} (1/z)=\pi \delta (\mathbf{r})$ and $[\del_z,\del_{\bar{z}}](1/z)=0$.   
A fundamental relation linking the vortex velocity field $v$ to the fluid velocity $u$ can also be derived by decomposing $J_{\rm n}$~\cite{SM1}, to give
\bea{\label{vortexvelocity}\rho v= \rho u -2i\eta \partial_z \sigma.}
The two velocity fields are related by 
\bea{\label{wvrelation}\rho w= \sigma v -i \eta \rho^{-1}\left(\partial_{z} \rho^2-\partial_z \sigma^2 \right),} 
and the vorticity of the vortex velocity field $v$, $\omega_v \equiv i(\partial_{\bar{z}} v-\partial_z \bar{v})$, has the anomalous correction 
\bea{\label{vorticityv}\omega_v-\omega=\nabla \times (\mathbf{v}-\mathbf{u})=\eta \nabla \cdot \left(\rho^{-1} \nabla \sigma\right) .}
In the hydrodynamic formulation the quantities $\rho$, $\sigma$, $J_{\rm c}$ and $J_{\rm n}$ represent averages of the corresponding microscopic quantities over patches, giving smooth coarse-grained quantities on the patch scale~\cite{SM2}.  

The relation~\eref{vortexvelocity} links the superfluid velocity field that is irregular at a vortex core, to a vortex fluid velocity field that is \emph{regular}. In other words, the velocity of a vortex at position $\mathbf{r}$ is the fluid velocity excluding the flow generated by the vortex itself at $\mathbf{r}$.  
The regularization involves subtracting the singular term, namely the pole at the vortex core. 
For a single vortex at the origin the superfluid velocity $u=-i\gam \sig_i/z$ and $2i\eta\partial_z\sig/\rho=4i\eta\sig_i/z$, Eq.~\eqref{vortexvelocity} yields  $v=u-2i\eta\partial_z\sig/\rho=0$.
The correction cancels out the superfluid velocity field due to the local vortex, giving the physical result that a vortex does not move under the action of its own velocity field.
Eq.~\eqref{canonical} has a similar interpretation.

The binary vortex fluid is compressible:
\bea{\label{compressibilityv}\nabla \cdot \mathbf{v}
	=2 \eta i \left[\partial_z (\rho^{-1} \partial_{\bar{z}} \sigma) - \rm {h.c.} \right]=-\eta \nabla \times (\rho^{-1}\nabla \sigma)\neq 0,} 
as also seen from $\nabla \cdot \mathbf{w}=\bar{u}\partial_{z}\left(\sigma/\rho\right)+u\partial_{\bar{z}}\left(\sigma/\rho\right)\neq 0$. 
In the chiral limit, $\sigma=\rho$, $w=v$ and $\nabla \cdot \mathbf{v}=0$~\cite{Wiegmann}. The chiral vortex fluid is rigid, as the energy cost to compress a vortex fluid containing $N$ vortices scales as $N^2$ due to repulsive interactions between like-sign vortices. For a binary vortex fluid, the presence of two opposite sign vortices softens the vortex fluid such that gapless excitations can occur, making the vortex fluid compressible.

\emph{Anomalous Euler Equation.|} Using Eqs.~\eqref{continuity1},~\eqref{wvrelation},~\eqref{compressibilityv}, $\sigma$ satisfies the vortex fluid Helmholtz equation
\bea{\label{Hwv}{\cal D}^{v}_t \sigma=0,}
and vortex charge is conserved, moving with the vortex velocity; use of Eq.~\eqref{vortexvelocity} shows consistency with Helmholtz' equation for the fluid, Eq.~\eqref{Hw}.  A  straightforward calculation~\cite{SM3} yields the anomalous Euler equation of the vortex velocity field $v$
\bea{\label{universalform}\partial_t(\rho v)+\partial_z {\cal T}_{z\bar z}+\partial_{\bar z} {\cal T}+\rho \partial_z(2p)=0,}
where 
${\cal T}_{z\bar z}=\rho v \bar v + 16 \eta^2 \pi \sigma^2+4 \eta^2\sigma \partial_{\bar z} (\rho^{-1}\partial_{z} \sigma)$ 
and	${\cal T} =\rho v v +4 \eta^2 \sigma \partial_z (\rho^{-1}\partial_z \sigma)-4\eta i \sigma \partial_z v$ are complex components of the momentum flux tensor.  
Eq.~\eqref{universalform} does not explicitly contain $w$ and  Eqs.~\eqref{continuity2}~\eqref{Hwv} and ~\eqref{universalform} fully describe the binary vortex fluid.
In Cartesian coordinates Eq.~\eqref{universalform} becomes 
\bea{\label{Cartesianuniversalform}\partial_t (\rho v_a)+ \partial_b {\cal T}_{ab}+\rho \partial_a p=0,}
where the momentum flux tensor
${\cal T}_{ab}=\rho v_a v_b - \Pi_{ab}$~\cite{LandauFluids},
with the emergent Cauchy stress tensor 
\bea{\Pi_{ab}=-\sigma \tau_{ab}-8\eta^2 \pi \sigma^2 \delta_{ab}- \eta^2 \sigma \partial_b(\rho^{-1}\partial_a \sigma),}
reflecting the macroscopic effects of the topological nature of quantum vortices.
The \emph{anomalous stress}
\bea{\tau_{xy}&=\tau_{yx}=\eta(\partial_x v_x-\partial_y v_y),\nn\\
	\tau_{xx}&=-\tau_{yy}=-\eta(\partial_x v_y+\partial_y v_x),}
does not cause energy dissipation and is formally identical to that of the chiral vortex fluid~\cite{Wiegmann}.
The anomalous stress $\tau_{ab}$ vanishes in uniform rotation with angular velocity $\Omega$, where $v=-\Omega i \bar z$. Although there is no frictional viscosity in the binary vortex fluid, the  shear stress ($\Pi_{ab} \neq 0$ for $a\neq b$) is non-vanishing, induced by gradients of $\rho$ and $\sigma$ and the anomalous stress $\tau_{ab}$. A similar situation can be found in the GPE when quantum pressure is important~\cite{vortexshedding}. 

A conspicuous feature of the Cauchy stress tensor $\Pi_{ab}$ for the  binary vortex fluid is that it is asymmetric, with non-trivial commutator linking to the compressibility
\bea{\label{asymmetric}
	\Pi_{xy}-\Pi_{yx}&=-\sigma \eta \nabla \cdot \mathbf{v}.}   
Note that under the transformation $x \leftrightarrow y$, the velocity $v_x \rightarrow -v_y$ and $v_y\rightarrow -v_x$,
and hence $\nabla \cdot \mathbf{v} \rightarrow -\nabla \cdot \mathbf{v}$.
For the chiral fluid $\sigma=\rho$, $\nabla \cdot \mathbf{v}=0$ and $\Pi_{ab}$ is symmetric~\cite{Wiegmann}.
The local mechanical pressure of the binary vortex fluid is given by the normal stress:
\bea{p_m&=-\frac{1}{2}{\rm tr} (\Pi_{ab})=8\eta^2 \pi \sigma^2 +\frac{1}{2}\eta^2 \sigma \nabla \cdot (\rho^{-1}\nabla \sigma).}

\emph{Angular Momentum.|}
The canonical angular momentum of the point-vortex system which is associated with rotational symmetry reads
$L^{\rm c}\equiv\sum^N_i \mathbf{r}_i \times \mathbf{p}_i=-\pi \gamma \sum_i \sigma_i r^2_i$, and is equivalent to the fluid angular momentum $L_{\rm f}=\int d^2\mathbf{r}\; \mathbf{r} \times \mathbf{u} 
= -1/2\int d^2\mathbf{r}\; r^2 \omega$. $L^{\rm c}$ is conserved as long as the system has rotational symmetry. We can also consider the orbital angular momentum (OAM) of vortices
$\label{orbitalAM}L^{\rm v} \equiv \sum^N_i \mathbf{r}_i \times \mathbf{v}_i$. For a chiral vortex system ($\sigma_i=1$), $L^{\rm v}=2 \eta N(N-1)$. In terms of the hydrodynamic field $v$,
$L^{\rm v}=\int d^2 \mathbf{r} {\cal L}^{\rm v}$ with ${\cal L}^{\rm v}=\mathbf {r} \times (\rho \mathbf{v})$.
In Cartesian coordinates ${\cal L}^{\rm v}_{ab}=\rho (x_a v_b-x_b v_a)$.
Using Eq.~\eqref{Cartesianuniversalform}, we obtain the continuity equation  
\bea{\label{OAMCE}\frac{\partial {\cal L}^{\rm v}_{ab}}{\partial t}+\partial_c {\cal M}_{abc}= \Pi_{ba}-\Pi_{ab}}
with angular momentum flux tensor 
${\cal M}_{abc}=x_a {\cal T}_{bc}-x_b {\cal T}_{ac}$, indicating that the OAM is conserved if and only if $\Pi_{ab}=\Pi_{ba}$. Since $\Pi_{ab}$ is asymmetric for the binary vortex fluid, the OAM is not conserved regardless of the symmetry, consistent with the property of the corresponding discrete point-vortex system ~\cite{SM4}. For the incompressible  chiral vortex fluid ($\nabla \cdot \mathbf{v}=0$), the OAM is conserved. The divergence
$\nabla \cdot \mathbf{v}$ is a source term in Eq.~\eqref{OAMCE}, and $\eta$ can be seen as a {\em rotational} viscous coefficient mediating the exchange between vortex fluid OAM and the internal vorticity and density degrees of freedom. The conserved OAM of the binary point-vortex system can be constructed by considering the sign-weighted OAM $L^{\rm w}\equiv \sum^N_i \mathbf{r}_i \times (\sigma_i \mathbf{v}_i)=2 \eta [(\sum_i \sigma_i)^2-N ]$.
In the chiral case $L^{\rm w}=L^{\rm v}$ and $L^{\rm w}=-2 \eta N$ for the neutral system; in terms of $w$,
$L^{\rm w}=\int d^2 \mathbf{r} \ \mathbf {r} \times (\rho \mathbf{w})$.

\emph{Vortex Fluid Hamiltonian.|}
Since $\sigma_i v_i \propto dp_i/dt$ and $\sum_i \delta(\mathbf{r}-\mathbf{r}_i)\partial {\cal H}/\partial z_i=\sum_i \delta(\mathbf{r}-\mathbf{r}_i) \sigma_i v_i$, the vortex charge velocity $w$ satisfies the canonical equation  
$-i\pi\gamma\rho w=\sigma \partial_z(\delta {\cal H }[\rho,\sigma]/\delta \sigma)+\rho \partial_z (\delta {\cal H }[\rho,\sigma]/\delta \rho)$, which gives the Hamiltonian of the binary vortex fluid~\cite{footnote7} 
\bea{\label{Hamiltoniancanonical}{\cal H}[\rho,\sigma]&=H[\sigma]-8\pi \eta^2 \int d^2 \mathbf{r}\;\rho \log (\ell^{2} \rho).}  
Here $H[\sigma]$ is the fluid Hamiltonian Eq.~\eqref{Hsf}, and the second term is the self-energy of vortices~\cite{footnote6}. 
The Hamiltonian Eq.~\eqref{Hamiltoniancanonical} is the hydrodynamic formulation to the discrete point-vortex Hamiltonian Eq.~\eqref{PointVortexH}. It can be  decomposed as ${\cal H}[\rho,\sigma]=H_{\rm kin}+H_{\rm int}+H_{\rm so}$, with vortex fluid
kinetic energy $H_{\rm kin}=1/2\int d^2\mathbf{r} |\mathbf{v}|^2$, ``internal energy" $H_{\rm int}=\eta^2/2\int d^2 \mathbf{r} (\rho^{-2}|\nabla \sigma|^2-16\pi  \rho \log \rho) $ and a ``spin-orbit coupling" term $H_{\rm so}=-\eta\int d^2\mathbf{r} \rho^{-1} \mathbf{v} \cdot (\hat{z} \times \nabla \sigma)$ that couples vortex distribution energy to vortex fluid kinetic energy.

\begin{figure}[!t]
	\includegraphics[width=3.2in]{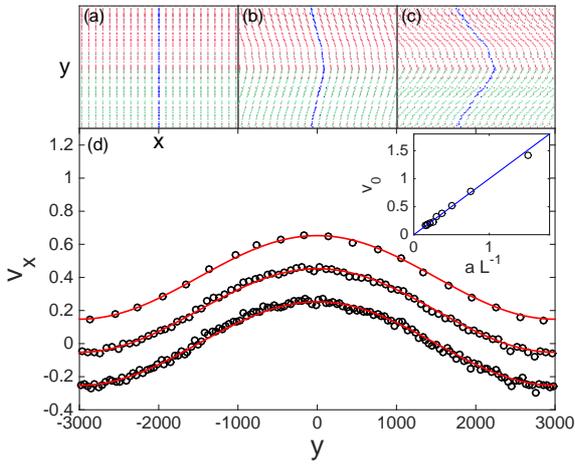}
	\caption{(Color online) The initial vortex configuration is obtained by sampling vortex coordinates according to the distribution $\sigma=\sigma_0 \sin(2 \pi y/L)$ with $y\in[-L/2,L/2]$, and $\sigma_0=\rho_0$. We measure length $L$, velocity $v$ and time $t$ in units of $\xi$, $\kappa/\xi$ and $\xi^2/\kappa$ respectively.
		(a), (b) and (c) show the evolution of point vortices (red: $\sigma_i=1$; green: $\sigma_i=-1$) at $t=0,50,150$ respectively. In each plot, $x$ and $y$ range from $-10^3$ to $10^3$ and $-3 \times 10^3$ to $3 \times 10^3$. The 138 vortices initially located at $x=0$ are labeled by blue and their velocities are used to characterize the shear flow. (d) Comparison between Eq.~\eqref{shearflow} (red) and  simulations (circles). Velocities of the blue colored vortices are obtained as averages over $10$ runs (bottom). By averaging the velocities over the increasing bin size $L/\{138,80,20\}$ (from bottom to top), the coarse-grained vortex velocities approach the analytical result~(\ref{shearflow}). For clarity the results for different bin sizes are vertically shifted by 0.2.
		The inset in (d) shows the estimated values (circles) of $v_0$ via fitting to the numerical data in a single run for each $L\in\{1, 2,...,10\}\times 10^3$, where $a=N/2\pi-1/4$; also shown is the analytical result (blue). In our sampling the constant density is anisotropic, with density along the $y$-axis twice that of the density along the $x$-axis.    
	} \label{shearflowFig}
\end{figure}

\emph{Vortex Shear Flows.|} We consider a static vortex flow with $v_y=0$, $\partial_x v_x=0$ and $\partial_x \sigma=0$ in a neutral vortex system of constant density $\rho=\rho_0$. For such a static flow  Eq.~\eqref{Cartesianuniversalform} reduces to $\partial_{b} {\cal T}_{ab}=\partial_{b} (\rho_0 v_a v_b- \Pi_{ab})=\partial_b \Pi_{ab}=0$~\cite{footnote14}, where the only non-trivial component reads
$\left(\partial_y \sigma \partial_y +\sigma \partial^2_y\right) v_x + \eta \left[16 \pi \sigma \partial_y +\rho^{-1}_0 (\partial_y \sigma \partial^2_y+\sigma \partial^3_y)\right]\sigma =0$,
indicating that $\mathbf{v}$ is completely determined by the emergent Cauchy stress tensor including the anomalous stress $\tau_{ab}$.  An exact shear flow solution to Eq.~\eqref{Cartesianuniversalform} can then be found: 
\bea{\label{shearflow}  \sigma(x,y)=\sigma_0\sin(\alpha y), \quad \mathbf{v}=(v_0 \cos(\alpha y),0),}
where $v_0=\alpha \eta \sigma_0 (8\pi \alpha^{-2}-\rho^{-1}_0)$, and $\alpha$ is a real parameter.
Numerical simulations of the PVM with $N=9522$~\cite{footnote17} in a doubly-periodic square box with side length $L=6\times 10^3\xi$~\cite{SM5} show excellent agreement with the prediction of Eq.~\eqref{shearflow} (see~\fref{shearflowFig}). 
Good agreement is also seen for $N=450$ and $L=300\xi$,
being nearly within reach of current BEC experiments~\cite{Relaxation2014Shin,Dalibard2015,Gauthier16}. 
 There may also be more exotic vortex flows with enhanced anomalous term in Eq.~\eqref{vortexvelocity}, as may be caused by a gradient discontinuity in $\sigma$. 

\emph{Dissipation and Annihilation.|} To model dissipation due to interaction between superfluid vortices and a non-condensed thermal cloud~\cite{Barenghi2009}, we consider a dissipative point-vortex model~\cite{dissipative1,dissipative2,Tfriction1,Tfriction2} of the form $d\bar{z}_i/dt= v_i+i \eta^{\ast} \sigma_i v_i$~\cite{footnote18}, 
where the dimensionless dissipation rate  $\eta^{\ast}$ measures energy damping~\cite{footnote16}. In the presence of dissipation, the continuity equations~\eqref{continuity2} and \eqref{continuity1} still hold subject to the substitutions $\mathbf{J}_n \rightarrow \mathbf{J}'_n=\mathbf{J}_n-\eta^{\ast} \mathbf{\hat{z}} \times \mathbf{J}_c $ and $\mathbf{J}_c \rightarrow \mathbf{J}'_c=\mathbf{J}_c-\eta^{\ast} \mathbf{\hat{z}} \times \mathbf{J}_n$. 
The dissipative Helmholtz equation is
${\cal D}^{\hat{v}}_t\sigma=-\eta^{\ast}\left(8\pi \eta \rho \sigma+\eta \nabla^2 \sigma-\mathbf{v} \times \nabla \rho\right)$,
where $\hat{\mathbf{v}}\equiv \mathbf{v}-\eta^{\ast}\eta \nabla \log \rho$, the terms $-8\pi \eta \eta^{\ast} \rho \sigma$ and $\eta^{\ast}\mathbf{v} \times \nabla \rho$ describe damping and transverse damping respectively. The negative sign in front of the diffusion term induces \emph {uphill diffusion} of $\sigma$, concentrating vorticity, and may contribute to inverse energy cascades and vortex clustering processes ~\cite{TomPRA,Angheluta17}. 

Various forms of vortex number loss can modify the Hamiltonian theory, including vortex annihilations due to collisions, and dissipative vortex dipole decay or boundary loss. For a low temperature BEC containing many vortices the dissipation rate is small, $\eta^{*}\simeq 10^{-4}$~\cite{AshtonPRX}, and annihilations due to collisions are the dominant limitation for the Hamiltonian PVM approach. However, their influence is regime-dependent~\cite{footnote13} and can be negligible for highly vorticity-polarized states, examples of which are the vortex shear flow (see ~\fref{shearflowFig})~\cite{footnote19}, the enstrophy cascade in decaying 2DQT~\cite{Matt2017} and Onsager clustered states~\cite{Matt2014,DecayingQTBillam,AnghelutaPRE}.  In the vortex dipole gas regime, dipole-dipole collisions are frequent and the theory is valid for times shorter than the dipole lifetime. The characteristic quantities are the collision cross-section, $\sigma_{\rm cs}\sim \xi$, the root-mean-square dipole velocity $v_{\rm {rms}} \sim \kappa d^{-1}$ (for $d \lesssim 10 \xi$ the Jones-Roberts soliton~\cite{jones1982motions,jones1986motions} forms), the mean free path ${\ell}_m \sim (\sigma_{\rm cs} \rho_{\rm d})^{-1}$, and the mean free time $\tau ={\ell}_m/ v_{\rm rms}\sim 10/(\kappa \rho_{\rm d})$, where $\rho_{\rm d}=\rho/2$ is the dipole density. The average total collision frequency per unit area is then $\tau^{-1}\rho_{\rm d}$, and assuming that every collision annihilates two vortices, we find the rate equation 
$dN/dt=-\Gamma_2 N^2$, where the two-body decay rate $\Gamma_2=2 \tau^{-1}\rho_{\rm d} A$ and $A$ is the system area.
Such a decay behavior has recently been reported~\cite{Relaxation2014Shin}; using their experimental parameters we estimate $\Gamma_2\simeq 0.054 \ s^{-1}$, close to the rate observed for the highest temperature. For the same parameters, the Hamiltonian PVM is a valid description up to the mean free time (lifetime) of a dipole, $\tau\sim 300{\rm ms}$~\cite{SM6}. For regimes at higher vortex energy than the dipole gas limit, the decay rate per vortex can be strongly suppressed~\cite{DecayingQTBillam}. Including vortex-antivortex annihilations into the hydrodynamic theory systematically is an important direction to explore in the future.

\emph{Conclusion.|} Our hydrodynamical formulation of the point-vortex model reveals that the collective motion of many vortices emerges as a binary vortex fluid with rich phenomenology including an asymmetric Cauchy stress tensor and compressible flow.
Examples of the former are relatively rare, and are associated with interactions between internal degrees of freedom and the bulk fluid flow, as occurs in liquid crystals~\cite{de1993physics}. Excellent agreement between the analytic solution for vortex shear flow and the numerical simulations demonstrates the validity of the coarse-grained approach to collective vortex motion. The compressible vortex fluid may support density waves, however global plane-waves appear as excitations on top of non-trivial static background flows and are thus likely to be transient in general. The hydrodynamic theory suggests many areas for exploration including local density and vorticity excitations, regular vortex flows, connections with classical flows, and states of fully-developed two-dimensional quantum turbulence.

\section*{Acknowledgments} 
We acknowledge M. T. Reeves, B. P. Anderson, P. B. Blakie, Giulio Giusteri, L. A. Toikka, L. Williamson, and J. M. Floryan for useful discussions. We are grateful to P. Wiegmann and A. G. Abanov for helpful feedback on an earlier version of the manuscript. ASB is supported by a Rutherford Discovery Fellowship administered by the Royal Society of New Zealand.

%\bibliography{References}

\begin{thebibliography}{96}%
	\makeatletter
	\providecommand \@ifxundefined [1]{%
		\@ifx{#1\undefined}
	}%
	\providecommand \@ifnum [1]{%
		\ifnum #1\expandafter \@firstoftwo
		\else \expandafter \@secondoftwo
		\fi
	}%
	\providecommand \@ifx [1]{%
		\ifx #1\expandafter \@firstoftwo
		\else \expandafter \@secondoftwo
		\fi
	}%
	\providecommand \natexlab [1]{#1}%
	\providecommand \enquote  [1]{``#1''}%
	\providecommand \bibnamefont  [1]{#1}%
	\providecommand \bibfnamefont [1]{#1}%
	\providecommand \citenamefont [1]{#1}%
	\providecommand \href@noop [0]{\@secondoftwo}%
	\providecommand \href [0]{\begingroup \@sanitize@url \@href}%
	\providecommand \@href[1]{\@@startlink{#1}\@@href}%
	\providecommand \@@href[1]{\endgroup#1\@@endlink}%
	\providecommand \@sanitize@url [0]{\catcode `\\12\catcode `\$12\catcode
		`\&12\catcode `\#12\catcode `\^12\catcode `\_12\catcode `\%12\relax}%
	\providecommand \@@startlink[1]{}%
	\providecommand \@@endlink[0]{}%
	\providecommand \url  [0]{\begingroup\@sanitize@url \@url }%
	\providecommand \@url [1]{\endgroup\@href {#1}{\urlprefix }}%
	\providecommand \urlprefix  [0]{URL }%
	\providecommand \Eprint [0]{\href }%
	\providecommand \doibase [0]{http://dx.doi.org/}%
	\providecommand \selectlanguage [0]{\@gobble}%
	\providecommand \bibinfo  [0]{\@secondoftwo}%
	\providecommand \bibfield  [0]{\@secondoftwo}%
	\providecommand \translation [1]{[#1]}%
	\providecommand \BibitemOpen [0]{}%
	\providecommand \bibitemStop [0]{}%
	\providecommand \bibitemNoStop [0]{.\EOS\space}%
	\providecommand \EOS [0]{\spacefactor3000\relax}%
	\providecommand \BibitemShut  [1]{\csname bibitem#1\endcsname}%
	\let\auto@bib@innerbib\@empty
	%</preamble>
	\bibitem [{\citenamefont {Kosterlitz}\ and\ \citenamefont
		{Thouless}(1972)}]{Thouless72}%
	\BibitemOpen
	\bibfield  {author} {\bibinfo {author} {\bibfnamefont {J.~M.}\ \bibnamefont
			{Kosterlitz}}\ and\ \bibinfo {author} {\bibfnamefont {D.~J.}\ \bibnamefont
			{Thouless}},\ }\href {http://stacks.iop.org/0022-3719/5/i=11/a=002}
	{\bibfield  {journal} {\bibinfo  {journal} {Journal of Physics C: Solid State
				Physics}\ }\textbf {\bibinfo {volume} {5}},\ \bibinfo {pages} {L124}
		(\bibinfo {year} {1972})}\BibitemShut {NoStop}%
	\bibitem [{\citenamefont {Kosterlitz}\ and\ \citenamefont
		{Thouless}(1973)}]{Thouless73}%
	\BibitemOpen
	\bibfield  {author} {\bibinfo {author} {\bibfnamefont {J.~M.}\ \bibnamefont
			{Kosterlitz}}\ and\ \bibinfo {author} {\bibfnamefont {D.~J.}\ \bibnamefont
			{Thouless}},\ }\href {http://stacks.iop.org/0022-3719/6/i=7/a=010} {\bibfield
		{journal} {\bibinfo  {journal} {Journal of Physics C: Solid State Physics}\
		}\textbf {\bibinfo {volume} {6}},\ \bibinfo {pages} {1181} (\bibinfo {year}
		{1973})}\BibitemShut {NoStop}%
	\bibitem [{\citenamefont {Bray}(1994)}]{Bray}%
	\BibitemOpen
	\bibfield  {author} {\bibinfo {author} {\bibfnamefont {A.}~\bibnamefont
			{Bray}},\ }\href {\doibase 10.1080/00018739400101505} {\bibfield  {journal}
		{\bibinfo  {journal} {Advances in Physics}\ }\textbf {\bibinfo {volume}
			{43}},\ \bibinfo {pages} {357} (\bibinfo {year} {1994})}\BibitemShut
	{NoStop}%
	\bibitem [{\citenamefont {Mondello}\ and\ \citenamefont
		{Goldenfeld}(1990)}]{Goldenfeld}%
	\BibitemOpen
	\bibfield  {author} {\bibinfo {author} {\bibfnamefont {M.}~\bibnamefont
			{Mondello}}\ and\ \bibinfo {author} {\bibfnamefont {N.}~\bibnamefont
			{Goldenfeld}},\ }\href {\doibase 10.1103/PhysRevA.42.5865} {\bibfield
		{journal} {\bibinfo  {journal} {Phys. Rev. A}\ }\textbf {\bibinfo {volume}
			{42}},\ \bibinfo {pages} {5865} (\bibinfo {year} {1990})}\BibitemShut
	{NoStop}%
	\bibitem [{\citenamefont {Bray}\ \emph {et~al.}(2000)\citenamefont {Bray},
		\citenamefont {Briant},\ and\ \citenamefont {Jervis}}]{Bray2000}%
	\BibitemOpen
	\bibfield  {author} {\bibinfo {author} {\bibfnamefont {A.~J.}\ \bibnamefont
			{Bray}}, \bibinfo {author} {\bibfnamefont {A.~J.}\ \bibnamefont {Briant}}, \
		and\ \bibinfo {author} {\bibfnamefont {D.~K.}\ \bibnamefont {Jervis}},\
	}\href {\doibase 10.1103/PhysRevLett.84.1503} {\bibfield  {journal} {\bibinfo
			{journal} {Phys. Rev. Lett.}\ }\textbf {\bibinfo {volume} {84}},\ \bibinfo
		{pages} {1503} (\bibinfo {year} {2000})}\BibitemShut {NoStop}%
	\bibitem [{\citenamefont {Barenghi}\ \emph {et~al.}(2001)\citenamefont
		{Barenghi}, \citenamefont {Donnelly},\ and\ \citenamefont
		{Vinen}}]{barenghi2001quantized}%
	\BibitemOpen
	\bibfield  {author} {\bibinfo {author} {\bibfnamefont {C.~F.}\ \bibnamefont
			{Barenghi}}, \bibinfo {author} {\bibfnamefont {R.~J.}\ \bibnamefont
			{Donnelly}}, \ and\ \bibinfo {author} {\bibfnamefont {W.}~\bibnamefont
			{Vinen}},\ }\href@noop {} {\emph {\bibinfo {title} {Quantized vortex dynamics
				and superfluid turbulence}}},\ Vol.\ \bibinfo {volume} {571}\ (\bibinfo
	{publisher} {Springer Science \& Business Media},\ \bibinfo {year}
	{2001})\BibitemShut {NoStop}%
	\bibitem [{\citenamefont {Vinen}(2006)}]{vinen2006introduction}%
	\BibitemOpen
	\bibfield  {author} {\bibinfo {author} {\bibfnamefont {W.}~\bibnamefont
			{Vinen}},\ }\href
	{https://link.springer.com/article/10.1007%2Fs10909-006-9240-6?LI=true}
		{\bibfield  {journal} {\bibinfo  {journal} {Journal of Low Temperature
					Physics}\ }\textbf {\bibinfo {volume} {145}},\ \bibinfo {pages} {7} (\bibinfo
			{year} {2006})}\BibitemShut {NoStop}%
		\bibitem [{\citenamefont {Barenghi}\ \emph {et~al.}(2014)\citenamefont
			{Barenghi}, \citenamefont {Skrbek},\ and\ \citenamefont
			{Sreenivasan}}]{barenghi2014introduction}%
		\BibitemOpen
		\bibfield  {author} {\bibinfo {author} {\bibfnamefont {C.~F.}\ \bibnamefont
				{Barenghi}}, \bibinfo {author} {\bibfnamefont {L.}~\bibnamefont {Skrbek}}, \
			and\ \bibinfo {author} {\bibfnamefont {K.~R.}\ \bibnamefont {Sreenivasan}},\
		}\href {http://www.pnas.org/content/111/Supplement_1/4647.abstract}
		{\bibfield  {journal} {\bibinfo  {journal} {Proceedings of the National
					Academy of Sciences}\ }\textbf {\bibinfo {volume} {111}},\ \bibinfo {pages}
			{4647} (\bibinfo {year} {2014})}\BibitemShut {NoStop}%
		\bibitem [{\citenamefont {Tsatsos}\ \emph {et~al.}(2016)\citenamefont
			{Tsatsos}, \citenamefont {Tavares}, \citenamefont {Cidrim}, \citenamefont
			{Fritsch}, \citenamefont {Caracanhas}, \citenamefont {dos Santos},
			\citenamefont {Barenghi},\ and\ \citenamefont {Bagnato}}]{Tsatsos:2016fu}%
		\BibitemOpen
		\bibfield  {author} {\bibinfo {author} {\bibfnamefont {M.~C.}\ \bibnamefont
				{Tsatsos}}, \bibinfo {author} {\bibfnamefont {P.~E.~S.}\ \bibnamefont
				{Tavares}}, \bibinfo {author} {\bibfnamefont {A.}~\bibnamefont {Cidrim}},
			\bibinfo {author} {\bibfnamefont {A.~R.}\ \bibnamefont {Fritsch}}, \bibinfo
			{author} {\bibfnamefont {M.~A.}\ \bibnamefont {Caracanhas}}, \bibinfo
			{author} {\bibfnamefont {F.~E.~A.}\ \bibnamefont {dos Santos}}, \bibinfo
			{author} {\bibfnamefont {C.~F.}\ \bibnamefont {Barenghi}}, \ and\ \bibinfo
			{author} {\bibfnamefont {V.~S.}\ \bibnamefont {Bagnato}},\ }\href
		{http://linkinghub.elsevier.com/retrieve/pii/S037015731600065X} {\bibfield
			{journal} {\bibinfo  {journal} {Phys. Rep.}\ }\textbf {\bibinfo {volume}
				{622}},\ \bibinfo {pages} {1} (\bibinfo {year} {2016})}\BibitemShut {NoStop}%
		\bibitem [{\citenamefont {Navon}\ \emph {et~al.}(2016)\citenamefont {Navon},
			\citenamefont {Gaunt}, \citenamefont {Smith},\ and\ \citenamefont
			{Hadzibabic}}]{Navon2016}%
		\BibitemOpen
		\bibfield  {author} {\bibinfo {author} {\bibfnamefont {N.}~\bibnamefont
				{Navon}}, \bibinfo {author} {\bibfnamefont {A.~L.}\ \bibnamefont {Gaunt}},
			\bibinfo {author} {\bibfnamefont {R.~P.}\ \bibnamefont {Smith}}, \ and\
			\bibinfo {author} {\bibfnamefont {Z.}~\bibnamefont {Hadzibabic}},\ }\href
		{\doibase 10.1038/nature20114} {\bibfield  {journal} {\bibinfo  {journal}
				{Nature}\ }\textbf {\bibinfo {volume} {539}},\ \bibinfo {pages} {72}
			(\bibinfo {year} {2016})}\BibitemShut {NoStop}%
		\bibitem [{\citenamefont {Anderson}(2016)}]{anderson2016fluid}%
		\BibitemOpen
		\bibfield  {author} {\bibinfo {author} {\bibfnamefont {B.~P.}\ \bibnamefont
				{Anderson}},\ }\href
		{https://www.nature.com/nature/journal/v539/n7627/abs/539036a.html}
		{\bibfield  {journal} {\bibinfo  {journal} {Nature}\ }\textbf {\bibinfo
				{volume} {539}},\ \bibinfo {pages} {36} (\bibinfo {year} {2016})}\BibitemShut
		{NoStop}%
		\bibitem [{\citenamefont {Tsubota}\ \emph {et~al.}(2017)\citenamefont
			{Tsubota}, \citenamefont {Fujimoto},\ and\ \citenamefont
			{Yui}}]{Tsubota2017}%
		\BibitemOpen
		\bibfield  {author} {\bibinfo {author} {\bibfnamefont {M.}~\bibnamefont
				{Tsubota}}, \bibinfo {author} {\bibfnamefont {K.}~\bibnamefont {Fujimoto}}, \
			and\ \bibinfo {author} {\bibfnamefont {S.}~\bibnamefont {Yui}},\ }\href
		{\doibase 10.1007/s10909-017-1789-8} {\bibfield  {journal} {\bibinfo
				{journal} {Journal of Low Temperature Physics}\ }\textbf {\bibinfo {volume}
				{188}},\ \bibinfo {pages} {119} (\bibinfo {year} {2017})}\BibitemShut
		{NoStop}%
		\bibitem [{\citenamefont {Henn}\ \emph {et~al.}(2009)\citenamefont {Henn},
			\citenamefont {Seman}, \citenamefont {Roati}, \citenamefont {Magalh{\~a}es},\
			and\ \citenamefont {Bagnato}}]{Henn09a}%
		\BibitemOpen
		\bibfield  {author} {\bibinfo {author} {\bibfnamefont {E.~A.~L.}\
				\bibnamefont {Henn}}, \bibinfo {author} {\bibfnamefont {J.~A.}\ \bibnamefont
				{Seman}}, \bibinfo {author} {\bibfnamefont {G.}~\bibnamefont {Roati}},
			\bibinfo {author} {\bibfnamefont {K.~M.~F.}\ \bibnamefont {Magalh{\~a}es}}, \
			and\ \bibinfo {author} {\bibfnamefont {V.~S.}\ \bibnamefont {Bagnato}},\
		}\href {http://link.aps.org/doi/10.1103/PhysRevLett.103.045301} {\bibfield
			{journal} {\bibinfo  {journal} {Phys. Rev. Lett.}\ }\textbf {\bibinfo
				{volume} {103}},\ \bibinfo {pages} {045301} (\bibinfo {year}
			{2009})}\BibitemShut {NoStop}%
		\bibitem [{foo({\natexlab{a}})}]{footnote0}%
		\BibitemOpen
		\href@noop {} {\emph {\bibinfo {title} {\rm{In a BEC, vortex bending can be
						significantly suppressed by trap confinement without destroying phase
						coherence, inducing effective two-dimensional vortex
						motion~\cite{NeelyPRL}.}}}}\BibitemShut {Stop}%
		\bibitem [{\citenamefont {Numasato}\ \emph {et~al.}(2010)\citenamefont
			{Numasato}, \citenamefont {Tsubota},\ and\ \citenamefont {L'vov}}]{DEC2010}%
		\BibitemOpen
		\bibfield  {author} {\bibinfo {author} {\bibfnamefont {R.}~\bibnamefont
				{Numasato}}, \bibinfo {author} {\bibfnamefont {M.}~\bibnamefont {Tsubota}}, \
			and\ \bibinfo {author} {\bibfnamefont {V.~S.}\ \bibnamefont {L'vov}},\ }\href
		{\doibase 10.1103/PhysRevA.81.063630} {\bibfield  {journal} {\bibinfo
				{journal} {Phys. Rev. A}\ }\textbf {\bibinfo {volume} {81}},\ \bibinfo
			{pages} {063630} (\bibinfo {year} {2010})}\BibitemShut {NoStop}%
		\bibitem [{\citenamefont {Neely}\ \emph {et~al.}(2013)\citenamefont {Neely},
			\citenamefont {Bradley}, \citenamefont {Samson}, \citenamefont {Rooney},
			\citenamefont {Wright}, \citenamefont {Law}, \citenamefont
			{Carretero-Gonz\'alez}, \citenamefont {Kevrekidis}, \citenamefont {Davis},\
			and\ \citenamefont {Anderson}}]{NeelyPRL}%
		\BibitemOpen
		\bibfield  {author} {\bibinfo {author} {\bibfnamefont {T.~W.}\ \bibnamefont
				{Neely}}, \bibinfo {author} {\bibfnamefont {A.~S.}\ \bibnamefont {Bradley}},
			\bibinfo {author} {\bibfnamefont {E.~C.}\ \bibnamefont {Samson}}, \bibinfo
			{author} {\bibfnamefont {S.~J.}\ \bibnamefont {Rooney}}, \bibinfo {author}
			{\bibfnamefont {E.~M.}\ \bibnamefont {Wright}}, \bibinfo {author}
			{\bibfnamefont {K.~J.~H.}\ \bibnamefont {Law}}, \bibinfo {author}
			{\bibfnamefont {R.}~\bibnamefont {Carretero-Gonz\'alez}}, \bibinfo {author}
			{\bibfnamefont {P.~G.}\ \bibnamefont {Kevrekidis}}, \bibinfo {author}
			{\bibfnamefont {M.~J.}\ \bibnamefont {Davis}}, \ and\ \bibinfo {author}
			{\bibfnamefont {B.~P.}\ \bibnamefont {Anderson}},\ }\href {\doibase
			10.1103/PhysRevLett.111.235301} {\bibfield  {journal} {\bibinfo  {journal}
				{Phys. Rev. Lett.}\ }\textbf {\bibinfo {volume} {111}},\ \bibinfo {pages}
			{235301} (\bibinfo {year} {2013})}\BibitemShut {NoStop}%
		\bibitem [{\citenamefont {Reeves}\ \emph {et~al.}(2012)\citenamefont {Reeves},
			\citenamefont {Anderson},\ and\ \citenamefont {Bradley}}]{Matt2012}%
		\BibitemOpen
		\bibfield  {author} {\bibinfo {author} {\bibfnamefont {M.~T.}\ \bibnamefont
				{Reeves}}, \bibinfo {author} {\bibfnamefont {B.~P.}\ \bibnamefont
				{Anderson}}, \ and\ \bibinfo {author} {\bibfnamefont {A.~S.}\ \bibnamefont
				{Bradley}},\ }\href {\doibase 10.1103/PhysRevA.86.053621} {\bibfield
			{journal} {\bibinfo  {journal} {Phys. Rev. A}\ }\textbf {\bibinfo {volume}
				{86}},\ \bibinfo {pages} {053621} (\bibinfo {year} {2012})}\BibitemShut
		{NoStop}%
		\bibitem [{\citenamefont {Bradley}\ and\ \citenamefont
			{Anderson}(2012)}]{AshtonPRX}%
		\BibitemOpen
		\bibfield  {author} {\bibinfo {author} {\bibfnamefont {A.~S.}\ \bibnamefont
				{Bradley}}\ and\ \bibinfo {author} {\bibfnamefont {B.~P.}\ \bibnamefont
				{Anderson}},\ }\href {\doibase 10.1103/PhysRevX.2.041001} {\bibfield
			{journal} {\bibinfo  {journal} {Phys. Rev. X}\ }\textbf {\bibinfo {volume}
				{2}},\ \bibinfo {pages} {041001} (\bibinfo {year} {2012})}\BibitemShut
		{NoStop}%
		\bibitem [{\citenamefont {Chesler}\ \emph {et~al.}(2013)\citenamefont
			{Chesler}, \citenamefont {Liu},\ and\ \citenamefont {Adams}}]{Chesler368}%
		\BibitemOpen
		\bibfield  {author} {\bibinfo {author} {\bibfnamefont {P.~M.}\ \bibnamefont
				{Chesler}}, \bibinfo {author} {\bibfnamefont {H.}~\bibnamefont {Liu}}, \ and\
			\bibinfo {author} {\bibfnamefont {A.}~\bibnamefont {Adams}},\ }\href
		{\doibase 10.1126/science.1233529} {\bibfield  {journal} {\bibinfo  {journal}
				{Science}\ }\textbf {\bibinfo {volume} {341}},\ \bibinfo {pages} {368}
			(\bibinfo {year} {2013})}\BibitemShut {NoStop}%
		\bibitem [{\citenamefont {Neely}\ \emph {et~al.}(2010)\citenamefont {Neely},
			\citenamefont {Samson}, \citenamefont {Bradley}, \citenamefont {Davis},\ and\
			\citenamefont {Anderson}}]{VortexDipolePRL}%
		\BibitemOpen
		\bibfield  {author} {\bibinfo {author} {\bibfnamefont {T.~W.}\ \bibnamefont
				{Neely}}, \bibinfo {author} {\bibfnamefont {E.~C.}\ \bibnamefont {Samson}},
			\bibinfo {author} {\bibfnamefont {A.~S.}\ \bibnamefont {Bradley}}, \bibinfo
			{author} {\bibfnamefont {M.~J.}\ \bibnamefont {Davis}}, \ and\ \bibinfo
			{author} {\bibfnamefont {B.~P.}\ \bibnamefont {Anderson}},\ }\href {\doibase
			10.1103/PhysRevLett.104.160401} {\bibfield  {journal} {\bibinfo  {journal}
				{Phys. Rev. Lett.}\ }\textbf {\bibinfo {volume} {104}},\ \bibinfo {pages}
			{160401} (\bibinfo {year} {2010})}\BibitemShut {NoStop}%
		\bibitem [{\citenamefont {Kwon}\ \emph {et~al.}(2014)\citenamefont {Kwon},
			\citenamefont {Moon}, \citenamefont {Choi}, \citenamefont {Seo},\ and\
			\citenamefont {Shin}}]{Relaxation2014Shin}%
		\BibitemOpen
		\bibfield  {author} {\bibinfo {author} {\bibfnamefont {W.~J.}\ \bibnamefont
				{Kwon}}, \bibinfo {author} {\bibfnamefont {G.}~\bibnamefont {Moon}}, \bibinfo
			{author} {\bibfnamefont {J.-y.}\ \bibnamefont {Choi}}, \bibinfo {author}
			{\bibfnamefont {S.~W.}\ \bibnamefont {Seo}}, \ and\ \bibinfo {author}
			{\bibfnamefont {Y.-i.}\ \bibnamefont {Shin}},\ }\href {\doibase
			10.1103/PhysRevA.90.063627} {\bibfield  {journal} {\bibinfo  {journal} {Phys.
					Rev. A}\ }\textbf {\bibinfo {volume} {90}},\ \bibinfo {pages} {063627}
			(\bibinfo {year} {2014})}\BibitemShut {NoStop}%
		\bibitem [{\citenamefont {Stagg}\ \emph {et~al.}(2015)\citenamefont {Stagg},
			\citenamefont {Allen}, \citenamefont {Parker},\ and\ \citenamefont
			{Barenghi}}]{Barenghi2015}%
		\BibitemOpen
		\bibfield  {author} {\bibinfo {author} {\bibfnamefont {G.~W.}\ \bibnamefont
				{Stagg}}, \bibinfo {author} {\bibfnamefont {A.~J.}\ \bibnamefont {Allen}},
			\bibinfo {author} {\bibfnamefont {N.~G.}\ \bibnamefont {Parker}}, \ and\
			\bibinfo {author} {\bibfnamefont {C.~F.}\ \bibnamefont {Barenghi}},\ }\href
		{\doibase 10.1103/PhysRevA.91.013612} {\bibfield  {journal} {\bibinfo
				{journal} {Phys. Rev. A}\ }\textbf {\bibinfo {volume} {91}},\ \bibinfo
			{pages} {013612} (\bibinfo {year} {2015})}\BibitemShut {NoStop}%
		\bibitem [{\citenamefont {Reeves}\ \emph {et~al.}(2013)\citenamefont {Reeves},
			\citenamefont {Billam}, \citenamefont {Anderson},\ and\ \citenamefont
			{Bradley}}]{IECMatt}%
		\BibitemOpen
		\bibfield  {author} {\bibinfo {author} {\bibfnamefont {M.~T.}\ \bibnamefont
				{Reeves}}, \bibinfo {author} {\bibfnamefont {T.~P.}\ \bibnamefont {Billam}},
			\bibinfo {author} {\bibfnamefont {B.~P.}\ \bibnamefont {Anderson}}, \ and\
			\bibinfo {author} {\bibfnamefont {A.~S.}\ \bibnamefont {Bradley}},\ }\href
		{\doibase 10.1103/PhysRevLett.110.104501} {\bibfield  {journal} {\bibinfo
				{journal} {Phys. Rev. Lett.}\ }\textbf {\bibinfo {volume} {110}},\ \bibinfo
			{pages} {104501} (\bibinfo {year} {2013})}\BibitemShut {NoStop}%
		\bibitem [{\citenamefont {Billam}\ \emph
			{et~al.}(2015{\natexlab{a}})\citenamefont {Billam}, \citenamefont {Reeves},\
			and\ \citenamefont {Bradley}}]{TomPRA}%
		\BibitemOpen
		\bibfield  {author} {\bibinfo {author} {\bibfnamefont {T.~P.}\ \bibnamefont
				{Billam}}, \bibinfo {author} {\bibfnamefont {M.~T.}\ \bibnamefont {Reeves}},
			\ and\ \bibinfo {author} {\bibfnamefont {A.~S.}\ \bibnamefont {Bradley}},\
		}\href {\doibase 10.1103/PhysRevA.91.023615} {\bibfield  {journal} {\bibinfo
				{journal} {Phys. Rev. A}\ }\textbf {\bibinfo {volume} {91}},\ \bibinfo
			{pages} {023615} (\bibinfo {year} {2015}{\natexlab{a}})}\BibitemShut
		{NoStop}%
		\bibitem [{\citenamefont {Skaugen}\ and\ \citenamefont
			{Angheluta}(2016)}]{AnghelutaPRE}%
		\BibitemOpen
		\bibfield  {author} {\bibinfo {author} {\bibfnamefont {A.}~\bibnamefont
				{Skaugen}}\ and\ \bibinfo {author} {\bibfnamefont {L.}~\bibnamefont
				{Angheluta}},\ }\href {\doibase 10.1103/PhysRevE.93.032106} {\bibfield
			{journal} {\bibinfo  {journal} {Phys. Rev. E}\ }\textbf {\bibinfo {volume}
				{93}},\ \bibinfo {pages} {032106} (\bibinfo {year} {2016})}\BibitemShut
		{NoStop}%
		\bibitem [{\citenamefont {Chomaz}\ \emph {et~al.}(2015)\citenamefont {Chomaz},
			\citenamefont {Corman}, \citenamefont {Bienaim{\'e}}, \citenamefont
			{Desbuquois}, \citenamefont {Weitenberg}, \citenamefont {Nascimb{\`e}ne},
			\citenamefont {Beugnon},\ and\ \citenamefont {Dalibard}}]{Dalibard2015}%
		\BibitemOpen
		\bibfield  {author} {\bibinfo {author} {\bibfnamefont {L.}~\bibnamefont
				{Chomaz}}, \bibinfo {author} {\bibfnamefont {L.}~\bibnamefont {Corman}},
			\bibinfo {author} {\bibfnamefont {T.}~\bibnamefont {Bienaim{\'e}}}, \bibinfo
			{author} {\bibfnamefont {R.}~\bibnamefont {Desbuquois}}, \bibinfo {author}
			{\bibfnamefont {C.}~\bibnamefont {Weitenberg}}, \bibinfo {author}
			{\bibfnamefont {S.}~\bibnamefont {Nascimb{\`e}ne}}, \bibinfo {author}
			{\bibfnamefont {J.}~\bibnamefont {Beugnon}}, \ and\ \bibinfo {author}
			{\bibfnamefont {J.}~\bibnamefont {Dalibard}},\ }\href
		{http://www.nature.com/articles/ncomms7162} {\bibfield  {journal} {\bibinfo
				{journal} {Nature communications}\ }\textbf {\bibinfo {volume} {6}} (\bibinfo
			{year} {2015})}\BibitemShut {NoStop}%
		\bibitem [{\citenamefont {Gauthier}\ \emph {et~al.}(2016)\citenamefont
			{Gauthier}, \citenamefont {Lenton}, \citenamefont {Parry}, \citenamefont
			{Baker}, \citenamefont {Davis}, \citenamefont {Rubinsztein-Dunlop},\ and\
			\citenamefont {Neely}}]{Gauthier16}%
		\BibitemOpen
		\bibfield  {author} {\bibinfo {author} {\bibfnamefont {G.}~\bibnamefont
				{Gauthier}}, \bibinfo {author} {\bibfnamefont {I.}~\bibnamefont {Lenton}},
			\bibinfo {author} {\bibfnamefont {N.~M.}\ \bibnamefont {Parry}}, \bibinfo
			{author} {\bibfnamefont {M.}~\bibnamefont {Baker}}, \bibinfo {author}
			{\bibfnamefont {M.~J.}\ \bibnamefont {Davis}}, \bibinfo {author}
			{\bibfnamefont {H.}~\bibnamefont {Rubinsztein-Dunlop}}, \ and\ \bibinfo
			{author} {\bibfnamefont {T.~W.}\ \bibnamefont {Neely}},\ }\href {\doibase
			10.1364/OPTICA.3.001136} {\bibfield  {journal} {\bibinfo  {journal}
				{\color{blue}Optica}\ }\textbf {\bibinfo {volume} {3}},\ \bibinfo {pages}
			{1136} (\bibinfo {year} {2016})}\BibitemShut {NoStop}%
		\bibitem [{\citenamefont {Seo}\ \emph {et~al.}(2017)\citenamefont {Seo},
			\citenamefont {Ko}, \citenamefont {Kim},\ and\ \citenamefont
			{Shin}}]{Shin2016}%
		\BibitemOpen
		\bibfield  {author} {\bibinfo {author} {\bibfnamefont {S.~W.}\ \bibnamefont
				{Seo}}, \bibinfo {author} {\bibfnamefont {B.}~\bibnamefont {Ko}}, \bibinfo
			{author} {\bibfnamefont {J.~H.}\ \bibnamefont {Kim}}, \ and\ \bibinfo
			{author} {\bibfnamefont {Y.}~\bibnamefont {Shin}},\ }\href {\doibase
			10.1038/s41598-017-04122-9} {\bibfield  {journal} {\bibinfo  {journal}
				{Scientific Reports}\ }\textbf {\bibinfo {volume} {7}},\ \bibinfo {pages}
			{4587} (\bibinfo {year} {2017})}\BibitemShut {NoStop}%
		\bibitem [{\citenamefont {Onsager}(1949)}]{Onsager1949}%
		\BibitemOpen
		\bibfield  {author} {\bibinfo {author} {\bibfnamefont {L.}~\bibnamefont
				{Onsager}},\ }\href {\doibase 10.1007/BF02780991} {\bibfield  {journal}
			{\bibinfo  {journal} {Il Nuovo Cimento (1943-1954)}\ }\textbf {\bibinfo
				{volume} {6}},\ \bibinfo {pages} {279} (\bibinfo {year} {1949})}\BibitemShut
		{NoStop}%
		\bibitem [{\citenamefont {Eyink}\ and\ \citenamefont
			{Sreenivasan}(2006)}]{EyinkRMP}%
		\BibitemOpen
		\bibfield  {author} {\bibinfo {author} {\bibfnamefont {G.~L.}\ \bibnamefont
				{Eyink}}\ and\ \bibinfo {author} {\bibfnamefont {K.~R.}\ \bibnamefont
				{Sreenivasan}},\ }\href {\doibase 10.1103/RevModPhys.78.87} {\bibfield
			{journal} {\bibinfo  {journal} {Rev. Mod. Phys.}\ }\textbf {\bibinfo {volume}
				{78}},\ \bibinfo {pages} {87} (\bibinfo {year} {2006})}\BibitemShut {NoStop}%
		\bibitem [{\citenamefont {Joyce}\ and\ \citenamefont
			{Montgomery}(1973)}]{JM73}%
		\BibitemOpen
		\bibfield  {author} {\bibinfo {author} {\bibfnamefont {G.}~\bibnamefont
				{Joyce}}\ and\ \bibinfo {author} {\bibfnamefont {D.}~\bibnamefont
				{Montgomery}},\ }\href {\doibase 10.1017/S0022377800007686} {\bibfield
			{journal} {\bibinfo  {journal} {Journal of Plasma Physics}\ }\textbf
			{\bibinfo {volume} {10}},\ \bibinfo {pages} {107} (\bibinfo {year}
			{1973})}\BibitemShut {NoStop}%
		\bibitem [{\citenamefont {Montgomery}\ and\ \citenamefont
			{Joyce}(1974)}]{MJ74}%
		\BibitemOpen
		\bibfield  {author} {\bibinfo {author} {\bibfnamefont {D.}~\bibnamefont
				{Montgomery}}\ and\ \bibinfo {author} {\bibfnamefont {G.}~\bibnamefont
				{Joyce}},\ }\href@noop {} {\bibfield  {journal} {\bibinfo  {journal} {Physics
					of Fluids}\ }\textbf {\bibinfo {volume} {17}} (\bibinfo {year}
			{1974})}\BibitemShut {NoStop}%
		\bibitem [{\citenamefont {Billam}\ \emph {et~al.}(2014)\citenamefont {Billam},
			\citenamefont {Reeves}, \citenamefont {Anderson},\ and\ \citenamefont
			{Bradley}}]{DecayingQTBillam}%
		\BibitemOpen
		\bibfield  {author} {\bibinfo {author} {\bibfnamefont {T.~P.}\ \bibnamefont
				{Billam}}, \bibinfo {author} {\bibfnamefont {M.~T.}\ \bibnamefont {Reeves}},
			\bibinfo {author} {\bibfnamefont {B.~P.}\ \bibnamefont {Anderson}}, \ and\
			\bibinfo {author} {\bibfnamefont {A.~S.}\ \bibnamefont {Bradley}},\ }\href
		{\doibase 10.1103/PhysRevLett.112.145301} {\bibfield  {journal} {\bibinfo
				{journal} {Phys. Rev. Lett.}\ }\textbf {\bibinfo {volume} {112}},\ \bibinfo
			{pages} {145301} (\bibinfo {year} {2014})}\BibitemShut {NoStop}%
		\bibitem [{\citenamefont {Simula}\ \emph {et~al.}(2014)\citenamefont {Simula},
			\citenamefont {Davis},\ and\ \citenamefont {Helmerson}}]{TapioPRL}%
		\BibitemOpen
		\bibfield  {author} {\bibinfo {author} {\bibfnamefont {T.}~\bibnamefont
				{Simula}}, \bibinfo {author} {\bibfnamefont {M.~J.}\ \bibnamefont {Davis}}, \
			and\ \bibinfo {author} {\bibfnamefont {K.}~\bibnamefont {Helmerson}},\ }\href
		{\doibase 10.1103/PhysRevLett.113.165302} {\bibfield  {journal} {\bibinfo
				{journal} {Phys. Rev. Lett.}\ }\textbf {\bibinfo {volume} {113}},\ \bibinfo
			{pages} {165302} (\bibinfo {year} {2014})}\BibitemShut {NoStop}%
		\bibitem [{\citenamefont {Yu}\ \emph {et~al.}(2016)\citenamefont {Yu},
			\citenamefont {Billam}, \citenamefont {Nian}, \citenamefont {Reeves},\ and\
			\citenamefont {Bradley}}]{clusteringYu}%
		\BibitemOpen
		\bibfield  {author} {\bibinfo {author} {\bibfnamefont {X.}~\bibnamefont
				{Yu}}, \bibinfo {author} {\bibfnamefont {T.~P.}\ \bibnamefont {Billam}},
			\bibinfo {author} {\bibfnamefont {J.}~\bibnamefont {Nian}}, \bibinfo {author}
			{\bibfnamefont {M.~T.}\ \bibnamefont {Reeves}}, \ and\ \bibinfo {author}
			{\bibfnamefont {A.~S.}\ \bibnamefont {Bradley}},\ }\href {\doibase
			10.1103/PhysRevA.94.023602} {\bibfield  {journal} {\bibinfo  {journal} {Phys.
					Rev. A}\ }\textbf {\bibinfo {volume} {94}},\ \bibinfo {pages} {023602}
			(\bibinfo {year} {2016})}\BibitemShut {NoStop}%
		\bibitem [{\citenamefont {Groszek}\ \emph {et~al.}(2016)\citenamefont
			{Groszek}, \citenamefont {Simula}, \citenamefont {Paganin},\ and\
			\citenamefont {Helmerson}}]{Groszek:2015ty}%
		\BibitemOpen
		\bibfield  {author} {\bibinfo {author} {\bibfnamefont {A.~J.}\ \bibnamefont
				{Groszek}}, \bibinfo {author} {\bibfnamefont {T.~P.}\ \bibnamefont {Simula}},
			\bibinfo {author} {\bibfnamefont {D.~M.}\ \bibnamefont {Paganin}}, \ and\
			\bibinfo {author} {\bibfnamefont {K.}~\bibnamefont {Helmerson}},\ }\href
		{http://link.aps.org/pdf/10.1103/PhysRevA.93.043614} {\bibfield  {journal}
			{\bibinfo  {journal} {Phys. Rev. A}\ }\textbf {\bibinfo {volume} {93}},\
			\bibinfo {pages} {043614} (\bibinfo {year} {2016})}\BibitemShut {NoStop}%
		\bibitem [{\citenamefont {Novikov}(1975)}]{novikov1975}%
		\BibitemOpen
		\bibfield  {author} {\bibinfo {author} {\bibfnamefont {E.}~\bibnamefont
				{Novikov}},\ }\href@noop {} {\bibfield  {journal} {\bibinfo  {journal}
				{Zhurnal Eksperimentalnoi i Teoreticheskoi Fiziki}\ }\textbf {\bibinfo
				{volume} {68}},\ \bibinfo {pages} {1868} (\bibinfo {year}
			{1975})}\BibitemShut {NoStop}%
		\bibitem [{\citenamefont {Aref}\ \emph {et~al.}(1999)\citenamefont {Aref},
			\citenamefont {Boyland}, \citenamefont {Stremler},\ and\ \citenamefont
			{Vainchtein}}]{Aref1999}%
		\BibitemOpen
		\bibfield  {author} {\bibinfo {author} {\bibfnamefont {H.}~\bibnamefont
				{Aref}}, \bibinfo {author} {\bibfnamefont {P.}~\bibnamefont {Boyland}},
			\bibinfo {author} {\bibfnamefont {M.}~\bibnamefont {Stremler}}, \ and\
			\bibinfo {author} {\bibfnamefont {D.}~\bibnamefont {Vainchtein}},\ }in\
		\href@noop {} {\emph {\bibinfo {booktitle} {Fundamental Problematic Issues in
					Turbulence}}}\ (\bibinfo  {publisher} {Springer},\ \bibinfo {year} {1999})\
		pp.\ \bibinfo {pages} {151--161}\BibitemShut {NoStop}%
		\bibitem [{\citenamefont {Fetter}(1967)}]{Fetter67}%
		\BibitemOpen
		\bibfield  {author} {\bibinfo {author} {\bibfnamefont {A.~L.}\ \bibnamefont
				{Fetter}},\ }\href {\doibase 10.1103/PhysRev.162.143} {\bibfield  {journal}
			{\bibinfo  {journal} {Phys. Rev.}\ }\textbf {\bibinfo {volume} {162}},\
			\bibinfo {pages} {143} (\bibinfo {year} {1967})}\BibitemShut {NoStop}%
		\bibitem [{\citenamefont {Haldane}\ and\ \citenamefont
			{Wu}(1985)}]{HaldaneYongshi}%
		\BibitemOpen
		\bibfield  {author} {\bibinfo {author} {\bibfnamefont {F.~D.~M.}\
				\bibnamefont {Haldane}}\ and\ \bibinfo {author} {\bibfnamefont {Y.-S.}\
				\bibnamefont {Wu}},\ }\href {\doibase 10.1103/PhysRevLett.55.2887} {\bibfield
			{journal} {\bibinfo  {journal} {Phys. Rev. Lett.}\ }\textbf {\bibinfo
				{volume} {55}},\ \bibinfo {pages} {2887} (\bibinfo {year}
			{1985})}\BibitemShut {NoStop}%
		\bibitem [{\citenamefont {Lucas}\ and\ \citenamefont
			{Sur\'owka}(2014)}]{Lucas}%
		\BibitemOpen
		\bibfield  {author} {\bibinfo {author} {\bibfnamefont {A.}~\bibnamefont
				{Lucas}}\ and\ \bibinfo {author} {\bibfnamefont {P.}~\bibnamefont
				{Sur\'owka}},\ }\href {\doibase 10.1103/PhysRevA.90.053617} {\bibfield
			{journal} {\bibinfo  {journal} {Phys. Rev. A}\ }\textbf {\bibinfo {volume}
				{90}},\ \bibinfo {pages} {053617} (\bibinfo {year} {2014})}\BibitemShut
		{NoStop}%
		\bibitem [{\citenamefont {Numasato}\ and\ \citenamefont
			{Tsubota}(2009)}]{Numasato2009}%
		\BibitemOpen
		\bibfield  {author} {\bibinfo {author} {\bibfnamefont {R.}~\bibnamefont
				{Numasato}}\ and\ \bibinfo {author} {\bibfnamefont {M.}~\bibnamefont
				{Tsubota}},\ }\href {\doibase 10.1007/s10909-009-9965-0} {\bibfield
			{journal} {\bibinfo  {journal} {Journal of Low Temperature Physics}\ }\textbf
			{\bibinfo {volume} {158}},\ \bibinfo {pages} {415} (\bibinfo {year}
			{2009})}\BibitemShut {NoStop}%
		\bibitem [{\citenamefont {Siggia}\ and\ \citenamefont
			{Aref}(1981)}]{siggia1981}%
		\BibitemOpen
		\bibfield  {author} {\bibinfo {author} {\bibfnamefont {E.~D.}\ \bibnamefont
				{Siggia}}\ and\ \bibinfo {author} {\bibfnamefont {H.}~\bibnamefont {Aref}},\
		}\href {\doibase 10.1063/1.863225} {\bibfield  {journal} {\bibinfo  {journal}
				{Physics of Fluids (1958-1988)}\ }\textbf {\bibinfo {volume} {24}},\ \bibinfo
			{pages} {171} (\bibinfo {year} {1981})}\BibitemShut {NoStop}%
		\bibitem [{\citenamefont {Skaugen}\ and\ \citenamefont
			{Angheluta}(2017)}]{Angheluta17}%
		\BibitemOpen
		\bibfield  {author} {\bibinfo {author} {\bibfnamefont {A.}~\bibnamefont
				{Skaugen}}\ and\ \bibinfo {author} {\bibfnamefont {L.}~\bibnamefont
				{Angheluta}},\ }\href {\doibase 10.1103/PhysRevE.95.052144} {\bibfield
			{journal} {\bibinfo  {journal} {Phys. Rev. E}\ }\textbf {\bibinfo {volume}
				{95}},\ \bibinfo {pages} {052144} (\bibinfo {year} {2017})}\BibitemShut
		{NoStop}%
		\bibitem [{\citenamefont {Reeves}\ \emph {et~al.}(2017)\citenamefont {Reeves},
			\citenamefont {Billam}, \citenamefont {Yu},\ and\ \citenamefont
			{Bradley}}]{Matt2017}%
		\BibitemOpen
		\bibfield  {author} {\bibinfo {author} {\bibfnamefont {M.~T.}\ \bibnamefont
				{Reeves}}, \bibinfo {author} {\bibfnamefont {T.~P.}\ \bibnamefont {Billam}},
			\bibinfo {author} {\bibfnamefont {X.}~\bibnamefont {Yu}}, \ and\ \bibinfo
			{author} {\bibfnamefont {A.~S.}\ \bibnamefont {Bradley}},\ }\href
		{https://arxiv.org/abs/1702.04445} {} (\bibinfo {year} {2017}),\ \Eprint
		{http://arxiv.org/abs/1702.04445} {arXiv:1702.04445} \BibitemShut {NoStop}%
		\bibitem [{\citenamefont {Wiegmann}\ and\ \citenamefont
			{Abanov}(2014)}]{Wiegmann}%
		\BibitemOpen
		\bibfield  {author} {\bibinfo {author} {\bibfnamefont {P.}~\bibnamefont
				{Wiegmann}}\ and\ \bibinfo {author} {\bibfnamefont {A.~G.}\ \bibnamefont
				{Abanov}},\ }\href {\doibase 10.1103/PhysRevLett.113.034501} {\bibfield
			{journal} {\bibinfo  {journal} {Phys. Rev. Lett.}\ }\textbf {\bibinfo
				{volume} {113}},\ \bibinfo {pages} {034501} (\bibinfo {year}
			{2014})}\BibitemShut {NoStop}%
		\bibitem [{\citenamefont {Smith}(1989)}]{SmithPRL}%
		\BibitemOpen
		\bibfield  {author} {\bibinfo {author} {\bibfnamefont {R.~A.}\ \bibnamefont
				{Smith}},\ }\href {\doibase 10.1103/PhysRevLett.63.1479} {\bibfield
			{journal} {\bibinfo  {journal} {Phys. Rev. Lett.}\ }\textbf {\bibinfo
				{volume} {63}},\ \bibinfo {pages} {1479} (\bibinfo {year}
			{1989})}\BibitemShut {NoStop}%
		\bibitem [{\citenamefont {Smith}\ and\ \citenamefont
			{O’Neil}(1990)}]{SmithONeil}%
		\BibitemOpen
		\bibfield  {author} {\bibinfo {author} {\bibfnamefont {R.~A.}\ \bibnamefont
				{Smith}}\ and\ \bibinfo {author} {\bibfnamefont {T.~M.}\ \bibnamefont
				{O’Neil}},\ }\href {http://aip.scitation.org/doi/abs/10.1063/1.859362}
		{\bibfield  {journal} {\bibinfo  {journal} {Physics of Fluids B}\ }\textbf
			{\bibinfo {volume} {2}},\ \bibinfo {pages} {2961} (\bibinfo {year}
			{1990})}\BibitemShut {NoStop}%
		\bibitem [{\citenamefont {Wiegmann}(2013)}]{VortexfluidHalleffect}%
		\BibitemOpen
		\bibfield  {author} {\bibinfo {author} {\bibfnamefont {P.~B.}\ \bibnamefont
				{Wiegmann}},\ }\href {\doibase 10.1103/PhysRevB.88.241305} {\bibfield
			{journal} {\bibinfo  {journal} {Phys. Rev. B}\ }\textbf {\bibinfo {volume}
				{88}},\ \bibinfo {pages} {241305} (\bibinfo {year} {2013})}\BibitemShut
		{NoStop}%
		\bibitem [{foo({\natexlab{b}})}]{footnote2}%
		\BibitemOpen
		\href@noop {} {\emph {\bibinfo {title} {\rm{We assume that the flow vanishes
						at the boundaries.}}}}\BibitemShut {Stop}%
		\bibitem [{foo({\natexlab{c}})}]{footnote11}%
		\BibitemOpen
		\href@noop {} {\emph {\bibinfo {title} {\rm{In this regime the Mach number
						$M\equiv|\mathbf{u}|/c \ll 1$, where $c$ is the speed of
						sound}}}}\BibitemShut {NoStop}%
		\bibitem [{\citenamefont {Pitaevskii}\ and\ \citenamefont
			{Stringari}(2016)}]{BECbook}%
		\BibitemOpen
		\bibfield  {author} {\bibinfo {author} {\bibfnamefont {L.}~\bibnamefont
				{Pitaevskii}}\ and\ \bibinfo {author} {\bibfnamefont {S.}~\bibnamefont
				{Stringari}},\ }\href@noop {} {\emph {\bibinfo {title} {Bose-Einstein
					Condensation and Superfluidity}}},\ Vol.\ \bibinfo {volume} {164}\ (\bibinfo
		{publisher} {Oxford University Press},\ \bibinfo {year} {2016})\BibitemShut
		{NoStop}%
		\bibitem [{\citenamefont {Kirchoff}(1877)}]{Kirchoff}%
		\BibitemOpen
		\bibfield  {author} {\bibinfo {author} {\bibfnamefont {G.}~\bibnamefont
				{Kirchoff}},\ }\href@noop {} {\emph {\bibinfo {title} {Lectures on
					Mathematical Physics, Mechanics}}}\ (\bibinfo  {publisher} {Teubner,
			Leipzig},\ \bibinfo {year} {1877})\BibitemShut {NoStop}%
		\bibitem [{\citenamefont {Chorin}(1994)}]{chorin}%
		\BibitemOpen
		\bibfield  {author} {\bibinfo {author} {\bibfnamefont {A.~J.}\ \bibnamefont
				{Chorin}},\ }\href@noop {} {\emph {\bibinfo {title} {Vorticity and
					turbulence}}},\ Vol.\ \bibinfo {volume} {103}\ (\bibinfo  {publisher}
		{Springer Science \& Business Media},\ \bibinfo {year} {1994})\BibitemShut
		{NoStop}%
		\bibitem [{\citenamefont {Marchioro}\ and\ \citenamefont
			{Pulvirenti}(2012)}]{mathvortex}%
		\BibitemOpen
		\bibfield  {author} {\bibinfo {author} {\bibfnamefont {C.}~\bibnamefont
				{Marchioro}}\ and\ \bibinfo {author} {\bibfnamefont {M.}~\bibnamefont
				{Pulvirenti}},\ }\href@noop {} {\emph {\bibinfo {title} {Mathematical theory
					of incompressible nonviscous fluids}}},\ Vol.~\bibinfo {volume} {96}\
		(\bibinfo  {publisher} {Springer Science \& Business Media},\ \bibinfo {year}
		{2012})\BibitemShut {NoStop}%
		\bibitem [{foo({\natexlab{d}})}]{footnote9}%
		\BibitemOpen
		\href@noop {} {\emph {\bibinfo {title} {\rm{This choice of density-weighted
						velocity fields is essential, avoiding the pathological velocity field where
						$\sigma=0$. For instance one may define $J_c=\sigma w'$ and $w'$ is
						ill-defined when $\sigma=0$. As the coordinates $(x,y)$ of a vortex are
						conjugate variables, Eq.~\eqref{continuity2} is recognised as the Liouville
						equation of the point-vortex system }}}}\BibitemShut {NoStop}%
		\bibitem [{SM1()}]{SM1}%
		\BibitemOpen
		\href@noop {} {\emph {\bibinfo {title} {\rm{See Supplemental Material for the
						detailed derivation.}}}}\BibitemShut {Stop}%
		\bibitem [{SM2()}]{SM2}%
		\BibitemOpen
		\href@noop {} {\emph {\bibinfo {title} {\rm{See Supplemental Material for the
						detailed explanation, which includes Ref.~\cite{Jackson}}.}}}\BibitemShut
		{Stop}%
		\bibitem [{SM3()}]{SM3}%
		\BibitemOpen
		\href@noop {} {\emph {\bibinfo {title} {\rm{See Supplemental Material for the
						detailed derivation}.}}}\BibitemShut {Stop}%
		\bibitem [{\citenamefont {Landau}\ and\ \citenamefont
			{Lifshitz}(2013)}]{LandauFluids}%
		\BibitemOpen
		\bibfield  {author} {\bibinfo {author} {\bibfnamefont {L.}~\bibnamefont
				{Landau}}\ and\ \bibinfo {author} {\bibfnamefont {E.}~\bibnamefont
				{Lifshitz}},\ }\href@noop {} {\emph {\bibinfo {title} {Fluid mechanics:
					Landau and Lifshitz: course of theoretical physics}}},\ Vol.~\bibinfo
		{volume} {6}\ (\bibinfo  {publisher} {Elsevier},\ \bibinfo {year}
		{2013})\BibitemShut {NoStop}%
		\bibitem [{\citenamefont {Winiecki}\ \emph {et~al.}(2000)\citenamefont
			{Winiecki}, \citenamefont {Jackson}, \citenamefont {McCann},\ and\
			\citenamefont {Adams}}]{vortexshedding}%
		\BibitemOpen
		\bibfield  {author} {\bibinfo {author} {\bibfnamefont {T.}~\bibnamefont
				{Winiecki}}, \bibinfo {author} {\bibfnamefont {B.}~\bibnamefont {Jackson}},
			\bibinfo {author} {\bibfnamefont {J.~F.}\ \bibnamefont {McCann}}, \ and\
			\bibinfo {author} {\bibfnamefont {C.~S.}\ \bibnamefont {Adams}},\ }\href
		{http://stacks.iop.org/0953-4075/33/i=19/a=317} {\bibfield  {journal}
			{\bibinfo  {journal} {Journal of Physics B: Atomic, Molecular and Optical
					Physics}\ }\textbf {\bibinfo {volume} {33}},\ \bibinfo {pages} {4069}
			(\bibinfo {year} {2000})}\BibitemShut {NoStop}%
		\bibitem [{SM4()}]{SM4}%
		\BibitemOpen
		\href@noop {} {\emph {\bibinfo {title} {\rm{See Supplemental Material for an
						example of a binary vortex system containing three vortices}.}}}\BibitemShut
		{Stop}%
		\bibitem [{foo({\natexlab{e}})}]{footnote7}%
		\BibitemOpen
		\href@noop {} {\emph {\bibinfo {title} {\rm{Note that the vortex velocity
						field $v$ is not a canonical variable and hence it does not have a simple
						canonical relation to the vortex fluid Hamiltonian.}}}}\BibitemShut {Stop}%
		\bibitem [{foo({\natexlab{f}})}]{footnote6}%
		\BibitemOpen
		\href@noop {} {\emph {\bibinfo {title} {\rm{Plugging
						$\rho=\sum_j\delta(\mathbf{r}-\mathbf{r}_j)$ into the last term of
						Eq.~\eqref{Hamiltoniancanonical}, we obtain $-8\pi \eta^2 \int d^2 \mathbf{r}
						\rho \log (\ell^2\rho)=-(1/2) \pi \gamma^2 \sum_i \log [\ell^2 \sum_j
						\delta(\mathbf{r}_i-\mathbf{r}_j)]\simeq -(N/2) \pi \gamma^2 \log
						(\ell^2/\xi^2)=N\pi \gamma^2 \log (\xi/\ell)=E_{\rm{self}}$ }}}}\BibitemShut
		{NoStop}%
		\bibitem [{foo({\natexlab{g}})}]{footnote14}%
		\BibitemOpen
		\href@noop {} {\emph {\bibinfo {title} {\rm{For such a flow, using
						Eq.~\eqref{vortexvelocity} one can show that $\nabla \cdot (\mathbf{u} \cdot
						\nabla \mathbf{u})=-\nabla^2 p=0$ , having a solution that solves $\partial_a
						p=0$}}}}\BibitemShut {NoStop}%
		\bibitem [{foo({\natexlab{h}})}]{footnote17}%
		\BibitemOpen
		\href@noop {} {\emph {\bibinfo {title} {\rm{A relatively large number of
						vortices ensures that a good statistics of the sampling can be
						achieved.}}}}\BibitemShut {Stop}%
		\bibitem [{SM5()}]{SM5}%
		\BibitemOpen
		\href@noop {} {\emph {\bibinfo {title} {\rm{See Supplemental Material for
						details of simulations, which includes Refs.~\cite{Pismen, doublyperiodic,
							Matt2017} }.}}}\BibitemShut {Stop}%
		\bibitem [{\citenamefont {Jackson}\ \emph {et~al.}(2009)\citenamefont
			{Jackson}, \citenamefont {Proukakis}, \citenamefont {Barenghi},\ and\
			\citenamefont {Zaremba}}]{Barenghi2009}%
		\BibitemOpen
		\bibfield  {author} {\bibinfo {author} {\bibfnamefont {B.}~\bibnamefont
				{Jackson}}, \bibinfo {author} {\bibfnamefont {N.~P.}\ \bibnamefont
				{Proukakis}}, \bibinfo {author} {\bibfnamefont {C.~F.}\ \bibnamefont
				{Barenghi}}, \ and\ \bibinfo {author} {\bibfnamefont {E.}~\bibnamefont
				{Zaremba}},\ }\href {\doibase 10.1103/PhysRevA.79.053615} {\bibfield
			{journal} {\bibinfo  {journal} {Phys. Rev. A}\ }\textbf {\bibinfo {volume}
				{79}},\ \bibinfo {pages} {053615} (\bibinfo {year} {2009})}\BibitemShut
		{NoStop}%
		\bibitem [{\citenamefont {Rica}\ and\ \citenamefont
			{Tirapegui}(1990)}]{dissipative1}%
		\BibitemOpen
		\bibfield  {author} {\bibinfo {author} {\bibfnamefont {S.}~\bibnamefont
				{Rica}}\ and\ \bibinfo {author} {\bibfnamefont {E.}~\bibnamefont
				{Tirapegui}},\ }\href {\doibase 10.1103/PhysRevLett.64.878} {\bibfield
			{journal} {\bibinfo  {journal} {Phys. Rev. Lett.}\ }\textbf {\bibinfo
				{volume} {64}},\ \bibinfo {pages} {878} (\bibinfo {year} {1990})}\BibitemShut
		{NoStop}%
		\bibitem [{\citenamefont {T\"ornkvist}\ and\ \citenamefont
			{Schr\"oder}(1997)}]{dissipative2}%
		\BibitemOpen
		\bibfield  {author} {\bibinfo {author} {\bibfnamefont {O.}~\bibnamefont
				{T\"ornkvist}}\ and\ \bibinfo {author} {\bibfnamefont {E.}~\bibnamefont
				{Schr\"oder}},\ }\href {\doibase 10.1103/PhysRevLett.78.1908} {\bibfield
			{journal} {\bibinfo  {journal} {Phys. Rev. Lett.}\ }\textbf {\bibinfo
				{volume} {78}},\ \bibinfo {pages} {1908} (\bibinfo {year}
			{1997})}\BibitemShut {NoStop}%
		\bibitem [{\citenamefont {Billam}\ \emph
			{et~al.}(2015{\natexlab{b}})\citenamefont {Billam}, \citenamefont {Reeves},\
			and\ \citenamefont {Bradley}}]{Tfriction1}%
		\BibitemOpen
		\bibfield  {author} {\bibinfo {author} {\bibfnamefont {T.~P.}\ \bibnamefont
				{Billam}}, \bibinfo {author} {\bibfnamefont {M.~T.}\ \bibnamefont {Reeves}},
			\ and\ \bibinfo {author} {\bibfnamefont {A.~S.}\ \bibnamefont {Bradley}},\
		}\href {\doibase 10.1103/PhysRevA.91.023615} {\bibfield  {journal} {\bibinfo
				{journal} {Phys. Rev. A}\ }\textbf {\bibinfo {volume} {91}},\ \bibinfo
			{pages} {023615} (\bibinfo {year} {2015}{\natexlab{b}})}\BibitemShut
		{NoStop}%
		\bibitem [{\citenamefont {Moon}\ \emph {et~al.}(2015)\citenamefont {Moon},
			\citenamefont {Kwon}, \citenamefont {Lee},\ and\ \citenamefont
			{Shin}}]{Tfriction2}%
		\BibitemOpen
		\bibfield  {author} {\bibinfo {author} {\bibfnamefont {G.}~\bibnamefont
				{Moon}}, \bibinfo {author} {\bibfnamefont {W.~J.}\ \bibnamefont {Kwon}},
			\bibinfo {author} {\bibfnamefont {H.}~\bibnamefont {Lee}}, \ and\ \bibinfo
			{author} {\bibfnamefont {Y.-i.}\ \bibnamefont {Shin}},\ }\href {\doibase
			10.1103/PhysRevA.92.051601} {\bibfield  {journal} {\bibinfo  {journal} {Phys.
					Rev. A}\ }\textbf {\bibinfo {volume} {92}},\ \bibinfo {pages} {051601}
			(\bibinfo {year} {2015})}\BibitemShut {NoStop}%
		\bibitem [{foo({\natexlab{i}})}]{footnote18}%
		\BibitemOpen
		\href@noop {} {\emph {\bibinfo {title} {\rm{Note that this velocity-dependent
						dissipation does not account for vortex-sound interactions, which are
						acceleration-dependent~\cite{Vinen,Barenghi2005}}}}}\BibitemShut {NoStop}%
		\bibitem [{foo({\natexlab{j}})}]{footnote16}%
		\BibitemOpen
		\href@noop {} {\emph {\bibinfo {title} {\rm{Such a damping term is inherited
						from the damped GPE~\cite{AshtonPRX} due to Bose-stimulated inelastic
						collisions between the condensate and non-condensed thermal cloud
		}}}}\BibitemShut {NoStop}%
		\bibitem [{foo({\natexlab{k}})}]{footnote13}%
		\BibitemOpen
		\href@noop {} {\emph {\bibinfo {title} {\rm{Including annihilations requires
						knowledge of vortex kinetics which are regime
						dependent~\cite{AHNS1980,Gasenzer2012pra,Gary2013,threebody2000,crossover2004,VS1991,TapioPRL,Relaxation2014Shin,
							Gasenzer2016}}}}}\BibitemShut {NoStop}%
		\bibitem [{foo({\natexlab{l}})}]{footnote19}%
		\BibitemOpen
		\href@noop {} {\emph {\bibinfo {title} {\rm{In our simulations, annihilations
						are modeled phenomenologically by removing vortex dipoles with separation
						less than $\xi$, and sound radiation of accelerating same-sign vortex pairs
						is modeled by smoothly increasing the dissipation $\eta^{*}$ as their
						separation decreases. Details can be found in the Supplemental Material.
						During our simulations only $3\%$ of vortices are lost }}}}\BibitemShut
		{NoStop}%
		\bibitem [{\citenamefont {Reeves}\ \emph {et~al.}(2014)\citenamefont {Reeves},
			\citenamefont {Billam}, \citenamefont {Anderson},\ and\ \citenamefont
			{Bradley}}]{Matt2014}%
		\BibitemOpen
		\bibfield  {author} {\bibinfo {author} {\bibfnamefont {M.~T.}\ \bibnamefont
				{Reeves}}, \bibinfo {author} {\bibfnamefont {T.~P.}\ \bibnamefont {Billam}},
			\bibinfo {author} {\bibfnamefont {B.~P.}\ \bibnamefont {Anderson}}, \ and\
			\bibinfo {author} {\bibfnamefont {A.~S.}\ \bibnamefont {Bradley}},\ }\href
		{\doibase 10.1103/PhysRevA.89.053631} {\bibfield  {journal} {\bibinfo
				{journal} {Phys. Rev. A}\ }\textbf {\bibinfo {volume} {89}},\ \bibinfo
			{pages} {053631} (\bibinfo {year} {2014})}\BibitemShut {NoStop}%
		\bibitem [{\citenamefont {Jones}\ and\ \citenamefont
			{Roberts}(1982)}]{jones1982motions}%
		\BibitemOpen
		\bibfield  {author} {\bibinfo {author} {\bibfnamefont {C.}~\bibnamefont
				{Jones}}\ and\ \bibinfo {author} {\bibfnamefont {P.}~\bibnamefont
				{Roberts}},\ }\href@noop {} {\bibfield  {journal} {\bibinfo  {journal}
				{Journal of Physics A: Mathematical and General}\ }\textbf {\bibinfo {volume}
				{15}},\ \bibinfo {pages} {2599} (\bibinfo {year} {1982})}\BibitemShut
		{NoStop}%
		\bibitem [{\citenamefont {Jones}\ \emph {et~al.}(1986)\citenamefont {Jones},
			\citenamefont {Putterman},\ and\ \citenamefont {Roberts}}]{jones1986motions}%
		\BibitemOpen
		\bibfield  {author} {\bibinfo {author} {\bibfnamefont {C.}~\bibnamefont
				{Jones}}, \bibinfo {author} {\bibfnamefont {S.}~\bibnamefont {Putterman}}, \
			and\ \bibinfo {author} {\bibfnamefont {P.}~\bibnamefont {Roberts}},\
		}\href@noop {} {\bibfield  {journal} {\bibinfo  {journal} {Journal of Physics
					A: Mathematical and General}\ }\textbf {\bibinfo {volume} {19}},\ \bibinfo
			{pages} {2991} (\bibinfo {year} {1986})}\BibitemShut {NoStop}%
		\bibitem [{SM6()}]{SM6}%
		\BibitemOpen
		\href@noop {} {\emph {\bibinfo {title} {\rm{See Supplemental Material for
						more detailed discussion, which includes Refs.~\cite{Gauthier16, Streed2006,
							Straten,Relaxation2014Shin,Gasenzer2012pra,Gasenzer2016,Barenghi2015,jones1982motions,jones1986motions,SM7}
					}.}}}\BibitemShut {Stop}%
		\bibitem [{\citenamefont {de~Gennes}\ and\ \citenamefont
			{Prost}(1995)}]{de1993physics}%
		\BibitemOpen
		\bibfield  {author} {\bibinfo {author} {\bibfnamefont {P.~G.}\ \bibnamefont
				{de~Gennes}}\ and\ \bibinfo {author} {\bibfnamefont {J.}~\bibnamefont
				{Prost}},\ }\href@noop {} {\enquote {\bibinfo {title} {The physics of liquid
					crystals},}\ } (\bibinfo {year} {1995})\BibitemShut {NoStop}%
		\bibitem [{\citenamefont {Jackson}(2007)}]{Jackson}%
		\BibitemOpen
		\bibfield  {author} {\bibinfo {author} {\bibfnamefont {J.~D.}\ \bibnamefont
				{Jackson}},\ }\href@noop {} {\emph {\bibinfo {title} {Classical
					electrodynamics}}}\ (\bibinfo  {publisher} {John Wiley \& Sons},\ \bibinfo
		{year} {2007})\BibitemShut {NoStop}%
		\bibitem [{\citenamefont {Pismen}(1999)}]{Pismen}%
		\BibitemOpen
		\bibfield  {author} {\bibinfo {author} {\bibfnamefont {L.}~\bibnamefont
				{Pismen}},\ }\href {http://books.google.co.nz/books?id=DtsZSOXCGaQC} {\emph
			{\bibinfo {title} {{Vortices in Nonlinear Fields}}}},\ International series
		of monographs on physics\ (\bibinfo  {publisher} {Clarendon Press},\ \bibinfo
		{address} {Oxford},\ \bibinfo {year} {1999})\BibitemShut {NoStop}%
		\bibitem [{\citenamefont {Weiss}\ and\ \citenamefont
			{McWilliams}(1991)}]{doublyperiodic}%
		\BibitemOpen
		\bibfield  {author} {\bibinfo {author} {\bibfnamefont {J.~B.}\ \bibnamefont
				{Weiss}}\ and\ \bibinfo {author} {\bibfnamefont {J.~C.}\ \bibnamefont
				{McWilliams}},\ }\href {http://dx.doi.org/10.1063/1.858014} {\bibfield
			{journal} {\bibinfo  {journal} {Physics of Fluids A: Fluid Dynamics}\
			}\textbf {\bibinfo {volume} {3}},\ \bibinfo {pages} {835} (\bibinfo {year}
			{1991})}\BibitemShut {NoStop}%
		\bibitem [{\citenamefont {Vinen}(2001)}]{Vinen}%
		\BibitemOpen
		\bibfield  {author} {\bibinfo {author} {\bibfnamefont {W.~F.}\ \bibnamefont
				{Vinen}},\ }\href {\doibase 10.1103/PhysRevB.64.134520} {\bibfield  {journal}
			{\bibinfo  {journal} {Phys. Rev. B}\ }\textbf {\bibinfo {volume} {64}},\
			\bibinfo {pages} {134520} (\bibinfo {year} {2001})}\BibitemShut {NoStop}%
		\bibitem [{\citenamefont {Barenghi}\ \emph {et~al.}(2005)\citenamefont
			{Barenghi}, \citenamefont {Parker}, \citenamefont {Proukakis},\ and\
			\citenamefont {Adams}}]{Barenghi2005}%
		\BibitemOpen
		\bibfield  {author} {\bibinfo {author} {\bibfnamefont {C.~F.}\ \bibnamefont
				{Barenghi}}, \bibinfo {author} {\bibfnamefont {N.}~\bibnamefont {Parker}},
			\bibinfo {author} {\bibfnamefont {N.}~\bibnamefont {Proukakis}}, \ and\
			\bibinfo {author} {\bibfnamefont {C.}~\bibnamefont {Adams}},\ }\href
		{\doibase 10.1007/s10909-005-2272-5} {\bibfield  {journal} {\bibinfo
				{journal} {Journal of Low Temperature Physics}\ }\textbf {\bibinfo {volume}
				{138}},\ \bibinfo {pages} {629} (\bibinfo {year} {2005})}\BibitemShut
		{NoStop}%
		\bibitem [{\citenamefont {Ambegaokar}\ \emph {et~al.}(1980)\citenamefont
			{Ambegaokar}, \citenamefont {Halperin}, \citenamefont {Nelson},\ and\
			\citenamefont {Siggia}}]{AHNS1980}%
		\BibitemOpen
		\bibfield  {author} {\bibinfo {author} {\bibfnamefont {V.}~\bibnamefont
				{Ambegaokar}}, \bibinfo {author} {\bibfnamefont {B.~I.}\ \bibnamefont
				{Halperin}}, \bibinfo {author} {\bibfnamefont {D.~R.}\ \bibnamefont
				{Nelson}}, \ and\ \bibinfo {author} {\bibfnamefont {E.~D.}\ \bibnamefont
				{Siggia}},\ }\href {\doibase 10.1103/PhysRevB.21.1806} {\bibfield  {journal}
			{\bibinfo  {journal} {Phys. Rev. B}\ }\textbf {\bibinfo {volume} {21}},\
			\bibinfo {pages} {1806} (\bibinfo {year} {1980})}\BibitemShut {NoStop}%
		\bibitem [{\citenamefont {Schole}\ \emph {et~al.}(2012)\citenamefont {Schole},
			\citenamefont {Nowak},\ and\ \citenamefont {Gasenzer}}]{Gasenzer2012pra}%
		\BibitemOpen
		\bibfield  {author} {\bibinfo {author} {\bibfnamefont {J.}~\bibnamefont
				{Schole}}, \bibinfo {author} {\bibfnamefont {B.}~\bibnamefont {Nowak}}, \
			and\ \bibinfo {author} {\bibfnamefont {T.}~\bibnamefont {Gasenzer}},\ }\href
		{\doibase 10.1103/PhysRevA.86.013624} {\bibfield  {journal} {\bibinfo
				{journal} {Phys. Rev. A}\ }\textbf {\bibinfo {volume} {86}},\ \bibinfo
			{pages} {013624} (\bibinfo {year} {2012})}\BibitemShut {NoStop}%
		\bibitem [{\citenamefont {Forrester}\ \emph {et~al.}(2013)\citenamefont
			{Forrester}, \citenamefont {Chu},\ and\ \citenamefont {Williams}}]{Gary2013}%
		\BibitemOpen
		\bibfield  {author} {\bibinfo {author} {\bibfnamefont {A.}~\bibnamefont
				{Forrester}}, \bibinfo {author} {\bibfnamefont {H.-C.}\ \bibnamefont {Chu}},
			\ and\ \bibinfo {author} {\bibfnamefont {G.~A.}\ \bibnamefont {Williams}},\
		}\href {\doibase 10.1103/PhysRevLett.110.165303} {\bibfield  {journal}
			{\bibinfo  {journal} {Phys. Rev. Lett.}\ }\textbf {\bibinfo {volume} {110}},\
			\bibinfo {pages} {165303} (\bibinfo {year} {2013})}\BibitemShut {NoStop}%
		\bibitem [{\citenamefont {Sire}\ and\ \citenamefont
			{Chavanis}(2000)}]{threebody2000}%
		\BibitemOpen
		\bibfield  {author} {\bibinfo {author} {\bibfnamefont {C.}~\bibnamefont
				{Sire}}\ and\ \bibinfo {author} {\bibfnamefont {P.-H.}\ \bibnamefont
				{Chavanis}},\ }\href {\doibase 10.1103/PhysRevE.61.6644} {\bibfield
			{journal} {\bibinfo  {journal} {Phys. Rev. E}\ }\textbf {\bibinfo {volume}
				{61}},\ \bibinfo {pages} {6644} (\bibinfo {year} {2000})}\BibitemShut
		{NoStop}%
		\bibitem [{\citenamefont {Yakhot}\ and\ \citenamefont
			{Wanderer}(2004)}]{crossover2004}%
		\BibitemOpen
		\bibfield  {author} {\bibinfo {author} {\bibfnamefont {V.}~\bibnamefont
				{Yakhot}}\ and\ \bibinfo {author} {\bibfnamefont {J.}~\bibnamefont
				{Wanderer}},\ }\href {\doibase 10.1103/PhysRevLett.93.154502} {\bibfield
			{journal} {\bibinfo  {journal} {Phys. Rev. Lett.}\ }\textbf {\bibinfo
				{volume} {93}},\ \bibinfo {pages} {154502} (\bibinfo {year}
			{2004})}\BibitemShut {NoStop}%
		\bibitem [{\citenamefont {Carnevale}\ \emph {et~al.}(1991)\citenamefont
			{Carnevale}, \citenamefont {McWilliams}, \citenamefont {Pomeau},
			\citenamefont {Weiss},\ and\ \citenamefont {Young}}]{VS1991}%
		\BibitemOpen
		\bibfield  {author} {\bibinfo {author} {\bibfnamefont {G.~F.}\ \bibnamefont
				{Carnevale}}, \bibinfo {author} {\bibfnamefont {J.~C.}\ \bibnamefont
				{McWilliams}}, \bibinfo {author} {\bibfnamefont {Y.}~\bibnamefont {Pomeau}},
			\bibinfo {author} {\bibfnamefont {J.~B.}\ \bibnamefont {Weiss}}, \ and\
			\bibinfo {author} {\bibfnamefont {W.~R.}\ \bibnamefont {Young}},\ }\href
		{\doibase 10.1103/PhysRevLett.66.2735} {\bibfield  {journal} {\bibinfo
				{journal} {Phys. Rev. Lett.}\ }\textbf {\bibinfo {volume} {66}},\ \bibinfo
			{pages} {2735} (\bibinfo {year} {1991})}\BibitemShut {NoStop}%
		\bibitem [{\citenamefont {Karl}\ and\ \citenamefont
			{Gasenzer}(2016)}]{Gasenzer2016}%
		\BibitemOpen
		\bibfield  {author} {\bibinfo {author} {\bibfnamefont {M.}~\bibnamefont
				{Karl}}\ and\ \bibinfo {author} {\bibfnamefont {T.}~\bibnamefont
				{Gasenzer}},\ }\href {https://arxiv.org/abs/1611.01163} {} (\bibinfo {year}
		{2016}),\ \Eprint {http://arxiv.org/abs/1611.01163} {arXiv:1611.01163}
		\BibitemShut {NoStop}%
		\bibitem [{\citenamefont {Streed}\ \emph {et~al.}(2006)\citenamefont {Streed},
			\citenamefont {Chikkatur}, \citenamefont {Gustavson}, \citenamefont {Boyd},
			\citenamefont {Torii}, \citenamefont {Schneble}, \citenamefont {Campbell},
			\citenamefont {Pritchard},\ and\ \citenamefont {Ketterle}}]{Streed2006}%
		\BibitemOpen
		\bibfield  {author} {\bibinfo {author} {\bibfnamefont {E.~W.}\ \bibnamefont
				{Streed}}, \bibinfo {author} {\bibfnamefont {A.~P.}\ \bibnamefont
				{Chikkatur}}, \bibinfo {author} {\bibfnamefont {T.~L.}\ \bibnamefont
				{Gustavson}}, \bibinfo {author} {\bibfnamefont {M.}~\bibnamefont {Boyd}},
			\bibinfo {author} {\bibfnamefont {Y.}~\bibnamefont {Torii}}, \bibinfo
			{author} {\bibfnamefont {D.}~\bibnamefont {Schneble}}, \bibinfo {author}
			{\bibfnamefont {G.~K.}\ \bibnamefont {Campbell}}, \bibinfo {author}
			{\bibfnamefont {D.~E.}\ \bibnamefont {Pritchard}}, \ and\ \bibinfo {author}
			{\bibfnamefont {W.}~\bibnamefont {Ketterle}},\ }\href
		{http://dx.doi.org/10.1063/1.2163977} {\bibfield  {journal} {\bibinfo
				{journal} {Review of Scientific Instruments}\ }\textbf {\bibinfo {volume}
				{77}},\ \bibinfo {pages} {023106} (\bibinfo {year} {2006})}\BibitemShut
		{NoStop}%
		\bibitem [{\citenamefont {van~der Stam}\ \emph {et~al.}(2007)\citenamefont
			{van~der Stam}, \citenamefont {van Ooijen}, \citenamefont {Meppelink},
			\citenamefont {Vogels},\ and\ \citenamefont {van~der Straten}}]{Straten}%
		\BibitemOpen
		\bibfield  {author} {\bibinfo {author} {\bibfnamefont {K.~M.~R.}\
				\bibnamefont {van~der Stam}}, \bibinfo {author} {\bibfnamefont {E.~D.}\
				\bibnamefont {van Ooijen}}, \bibinfo {author} {\bibfnamefont
				{R.}~\bibnamefont {Meppelink}}, \bibinfo {author} {\bibfnamefont {J.~M.}\
				\bibnamefont {Vogels}}, \ and\ \bibinfo {author} {\bibfnamefont
				{P.}~\bibnamefont {van~der Straten}},\ }\href
		{http://dx.doi.org/10.1063/1.2424439} {\bibfield  {journal} {\bibinfo
				{journal} {Review of Scientific Instruments}\ }\textbf {\bibinfo {volume}
				{78}},\ \bibinfo {pages} {013102} (\bibinfo {year} {2007})}\BibitemShut
		{NoStop}%
		\bibitem [{SM7()}]{SM7}%
		\BibitemOpen
		\href@noop {} {\emph {\bibinfo {title} {\rm{See Supplemental Material at
						http://link.aps.org/ supplemental/10.1103/PhysRevLett.111.235301 for
						additional information.}}}}\BibitemShut {Stop}%
	\end{thebibliography}



\section{Supplemental Material}

%\begin{abstract}
%In this supplementary material we provide details of constructions of hydrodynamic velocities and demonstrations on that the orbital angular momentum of a binary point vortex system is not conserved for the manuscript: ``Emergent Non-Eulerian Hydrodynamics of Quantum Vortices''.
%\end{abstract}

\maketitle

\subsection{Hydrodynamic velocities}
We first demonstrate Eq.~\eqref{canonical} in the main manuscript. Combining Eqs.~\eqref{Kirchhoff},~\eqref{micJ},~\eqref{decompozationbinary} in the main manuscript, we obtain   
\bea{\rho w &= \sum^{N}_{i=1} \delta(\mathbf{r}-\mathbf{r}_i)(\sigma_i v_i)\nn\\
	&=\frac{1}{\pi}\partial_{\bar{z}}\sum^{N}_{i=1}\frac{\sigma_i}{z-z_i}\sum_{j,j\neq i}\frac{-i\gamma \sigma_j}{z_i-z_j}\nn\\
	&=\frac{-i\gamma}{2\pi}\partial_{\bar z}\left[\left(\sum^{N}_{i=1} \frac{\sigma_i}{z-z_i}\right)^2+\partial_{z}\sum^{N}_{i=1}\frac{1}{z-z_i} \right] \nn \\
	&= \sigma u-2\eta i\partial_z \rho,}
where $\eta=\gamma/4$. In the last step we have used $\partial_{\bar z} (1/z)=\pi \delta (\mathbf{r})$ and $[\del_z,\del_{\bar{z}}](1/z)=0$.

Let us now consider the relation between the vortex velocity field $v$ and the fluid velocity $u$. For this aim we introduce  
\bea{\rho_{+}\equiv \sum^{N_+}_{i=1} \delta(\mathbf{r}-\mathbf{r}^{+}_i), \quad \text{and} \quad \rho_{-}\equiv \sum^{N_-}_{i=1} \delta(\mathbf{r}-\mathbf{r}^{-}_i),}
satisfying $\sigma=\rho_+ -\rho_- $ and  $\rho=\rho_+ + \rho_-$. Here $\mathbf{r}^{+}_i$ is the position of a positive vortex and $\mathbf{r}^{-}_i$ is the position of an anti-vortex.

The fluid velocity has a decomposition:    
\bea{\label{singularneutral}u=-\sum^N_{i=1}\frac{i \gamma \sigma_i }{z-z_i(t)}=u^{+}+u^{-},}
where
\bea{u^{+}=-\sum^{N_{+}}_{i=1}\frac{i \gamma \sigma^{+}_i }{z-z^{+}_i(t)} \quad \text{and} \quad u^{-}=-\sum^{N_{-}}_{i=1}\frac{i \gamma \sigma^{-}_i }{z-z^{-}_i(t)} }
are the fluid velocities generated by vortices with positive circulation and with negative circulation respectively. Here $\sigma^{\pm}_i=\pm1$.

Introducing the velocity of a vortex with positive (negative)
circulation
\bea{v^{\pm}_i \equiv-\sum^N_{j,j\neq i} \frac{i \gamma \sigma_j}{z^{\pm}_i(t)-z_j(t)},}
we can write 
\bea{\rho v=\sum^{N}_{i=1}\delta(\mathbf{r}-\mathbf{r}_i)v_i= \frac{1}{\pi}\partial_{\bar{z}}\sum^{N}_{i=1}\frac{v_i}{z-z_i}=\cal{I}_++ \cal{I}_-}
where 
\bea{{\cal I}_+=\frac{1}{\pi}\partial_{\bar{z}}\sum^{N_+}_{i=1}\frac{v^{+}_i}{z-z^{+}_i} \quad \text{and} \quad {\cal I}_-=\frac{1}{\pi}\partial_{\bar{z}}\sum^{N_-}_{i=1}\frac{v^{-}_i}{z-z^{-}_i}.}
We can further decompose ${\cal I}_{+}$ into 
\bea{{\cal I}_+={\cal I}^{+}_++{\cal I}^{-}_+, }
where
\bea{{\cal I}^{+}_+&=\frac{1}{\pi}\partial_{\bar{z}}\sum^{N_+}_{i=1}\frac{1}{z-z^{+}_i}\sum^{N_+}_{j,j\neq i}\frac{-i\gamma \sigma^{+}_j }{z^{+}_i-z^{+}_j}\nn\\
	&=\frac{-i\gamma}{2\pi}\partial_{\bar z}\left[\left(\sum^{N_+}_{i=1} \frac{1}{z-z^{+}_i}\right)^2+\partial_{z}\sum^{N_+}_{i=1}\frac{1}{z-z^{+}_i} \right] \nn\\
	&= \rho_+ u^{+} -\frac{i\gamma}{2}\partial_z \rho_{+},}
and
\bea{\label{flowvelocity1}{\cal I}^{-}_+&=\frac{1}{\pi}\partial_{\bar{z}}\sum^{N_+}_{i=1}\frac{1}{z-z^{+}_i}\sum^{N_-}_{j=1}\frac{-i\gamma \sigma^{-}_j}{z^{+}_i-z^{-}_j}\nn\\
	&=\sum^{N_+}_{i=1}\delta(\mathbf{r}-\mathbf{r}^{+}_i)u^{-}(z^{+}_i)\nn\\
	&=\rho_+ u^{-}.}
In order to see the last step of Eq.~\eqref{flowvelocity1} is true, let us integrate Eq.~\eqref{flowvelocity1} over an arbitrary region $\Omega$ which contains $M$ positive vortices. On one hand, we obtain
\bea{\int_{\Omega} d^2\mathbf{r} \sum^{N_+}_{i=1}\delta(\mathbf{r}-\mathbf{r}^{+}_i)u^{-}(z^{+}_i)&=\sum^{M}_{i=1} u^{-}(z^{+}_i).}
On the other hand side, we obtain 
\bea{\int_{\Omega} d^2\mathbf{r} \rho_{+} u^{-}&= \sum^{N_-}_{j=1}\int_{\Omega} d^2\mathbf{r}  \sum^{N_+}_{i=1}\delta(\mathbf{r}-\mathbf{r}^{+}_i) \frac{-i\sigma^{-}_j}{z-z^{-}_j}\nn\\
	&=\sum^{M}_{i=1} u^{-}(z^{+}_i), }
showing that Eq.~\eqref{flowvelocity1} is correct.

Similarly for ${\cal I}_{-}$ we have 
\bea{{\cal I}_-
	&=\frac{1}{\pi}\partial_{\bar{z}}\sum^{N_-}_{i=1}\frac{1}{z-z^{-}_i}\sum^{N_+}_{j,j\neq i}\frac{-i\gamma \sigma^{+}_j}{z^{-}_i-z^{+}_j} \nn\\
	&+\frac{1}{\pi}\partial_{\bar{z}}\sum^{N_-}_{i=1}\frac{1}{z-z^{-}_{i}}\sum^{N_-}_{j,j\neq i}\frac{-i\gamma \sigma^{-}_j}{z^{-}_i-z^{-}_j}\nn\\
	&=\rho_-u^{-}+\frac{i\gamma}{2}\partial_z \rho_-+\rho_- u^{+}.}
Collecting all the terms above we finally obtain Eq.~\eqref{vortexvelocity} in the main manuscript:
\bea{\rho v= \rho u -2i\eta \partial_z \sigma.} 

\subsection{Orbital Angular Momentum of Point Vortex Systems}~\label{orbital}
We show that the orbital angular momentum is not conserved for a binary vortex system with $N=3$. Firstly, let us consider $N=2$ with $\sigma_1=-\sigma_2=1$. The velocities of the two vortices read
\bea{\frac{d \bar z_1}{d t}=v_1=\frac{i\gamma}{z_1-z_2}, \quad \frac{d \bar z_2}{d t}=v_2=-\frac{i\gamma}{z_2-z_1}=v_1.}
The orbital angular momentum of the two opposite vortices is
\bea{L^{\rm v}= \frac{1}{2i} \sum_j (\bar z_j \bar v_j - z_j v_j)=-\frac{\gamma}{2} \left( \frac{\bar z_1 + \bar z_2}{\bar z_1 -\bar z_2}+ \frac{z_1 + z_2}{z_1-z_2} \right),}
and then 
\bea{\frac{d L^{\rm v}}{d t}=-\frac{\gamma^2}{16}\left(\frac{i}{|z_1-z_2|^2}+\rm h.c.\right)=0.}

We now consider $N=3$ with $\sigma_1=\sigma_2=-\sigma_3=1$, and then 
\bea{\frac{d L^{\rm v}}{d t}&=\frac{1}{2i}\sum_j\left( \frac{ d\bar z_j}{d t} \bar v_j + \frac{d \bar v_j}{d t} \bar z_j-\rm h.c.\right) \nn\\
	&=\frac{1}{2i}\sum_j\left(|v_j|^2 + \frac{d \bar v_j}{d t} \bar z_j-\rm h.c.\right) \nn\\
	&=\frac{1}{2i}\sum_j\left(\frac{d \bar v_j}{d t} \bar z_j-\rm h.c.\right).}
It is straightforward to find
\bea{\sum^3_{j=1} \frac{d \bar v_j}{d t} \bar z_j= - \gamma^2 \left[\frac{(\bar z_1 -\bar z_2)(z_1-z_2)}{(\bar z_1 -\bar z_2)^2 (z_2 -z_3)(z_1-z_3)} \right. \nn\\
	\left.+\frac{(\bar z_1 +\bar z_3)(z_1-z_3)}{(\bar z_1 -\bar z_3)^2 (z_3 -z_2)(z_1-z_2)}+ \right. \nn\\
	\left. \frac{(\bar z_2 +\bar z_3)(z_2-z_3)}{(\bar z_2 -\bar z_3)^2 (z_3 -z_1)(z_2-z_1)}\right], }
and in general 
\bea{\Im \left(\sum^3_{j=1} \frac{d \bar v_j}{d t} \bar z_j\right) \neq 0.}
For example for $z_1=1$, $z_2=2+i$ and $z_3=1+2 i$, $\Im \left(\sum^3_{j=1} \frac{d \bar v_j}{d t} \bar z_j\right)=1$.
Therefore, in general 
\bea{\label{s}\frac{d L^{\rm v}}{d t} \neq 0.}
It is expected that Eq.~\eqref{s} holds for a binary point vortex system with $N>3$. This is consistent with the corresponding coarse-grained vortex flow, where the Cauchy stress tensor is asymmetric.

\subsection{Derivation of the Anomalous Euler equation}
Here we outline the derivation of Eq.~\eqref{universalform} in the main manuscript. Combining the Euler equation   
\bea{\partial_t u+\partial_{\bar{z}}u^2+\partial_z ( \bar{u}u)+\partial_z(2p)=0,}
the continuity equation 
\bea{\partial_t \rho+[\partial_z (\rho \bar{v})+\partial_{\bar{z}} (\rho v)]=0, \label{continue}}
the Helmholtz equation 
\bea{{\cal D}^{v}_t \sigma=0, \label{HH}}
and the relation 
\bea{\rho v=\rho u -2i\eta \partial_z \sigma,	\label{relation} }
we obtain 
\bea{\label{EulerI}&\partial_t (\rho v)+\partial_{\bar{z}} (\rho v v)+\partial_{z} (\rho v \bar{v})= \\ 
	&-\rho \partial_z(2p) +u \partial_t \rho -2\eta i \partial_z \partial_t \sigma+u^2 \partial_{\bar{z}}\rho  -2i \eta \partial_{\bar{z}} u \partial_{z} \sigma \nn\\
	&-4 i\eta u \partial_{\bar{z}} \partial_z \sigma 
	-4 \eta^2 \partial_{\bar{z}} [\rho^{-1}(\partial_z \sigma)^2]
	+\bar{u}u \partial_z \rho + 2i \eta \partial_z u \partial_{\bar{z}} \sigma \nn\\
	&+2i \eta u \partial_z \partial_{\bar{z}} \sigma -2i \eta \bar{u} \partial_z \partial_z \sigma + 4 \eta^2 \partial_z (\rho^{-1}\partial_z \sigma \partial_{\bar{z}} \sigma). \nn}
This can be simplified as follows. Using Eqs.~\eqref{continue} and ~\eqref{HH}, we have  
\bea{\label{t1}u \partial_t \rho=&-(v+2i\eta \rho^{-1} \partial_z \sigma) [\partial_z (\rho \bar{v})+\partial_{\bar{z}} (\rho v)] \\
	=&-v\bar{v} \partial_z \rho -\rho v \partial_z \bar{v} -v^2 \partial_{\bar{z}} \rho -\rho v \partial_{\bar{z}} v 
	-2i \eta  \bar{v} \rho^{-1} \partial_z \rho \partial_z \sigma \nn\\
	&-2i \eta \partial_z \bar{v} \partial_z \sigma -2 \eta i \rho^{-1} v \partial_{\bar{z}} \rho \partial_z \sigma -2 \eta i \partial_{\bar{z}} v \partial_z \sigma, \nn}
and
\bea{\label{t2}-2\eta i \partial_z \partial_t \sigma
	&=2\eta i \partial_z \sigma \partial_z \bar{v}+ 2\eta  i \sigma \partial_z\partial_z \bar{v}+2\eta i \partial_z \bar{v} \partial_z \sigma +2\eta i \bar{v} \partial_z \partial_z \sigma \nn\\
	&+ 2 \eta i \partial_{\bar{z}} (\sigma \partial_z v) + 2\eta i v \partial_z \partial_{\bar{z}} \sigma +2\eta i \partial_z \sigma \partial_{\bar{z}} v \nn \\
	&+2 \eta^2 \partial_z \left(\partial_z\rho^{-1}\partial_{\bar{z}}\sigma^2-\partial_{\bar{z}} \rho^{-1}\partial_{z}\sigma^2 \right).}
Substituting Eqs.~\eqref{t1} and ~\eqref{t2} in to Eq.~\eqref{EulerI} and expressing $u$ in terms of $v$ and $\sigma$ using Eq.~\eqref{relation}, many terms on the right hand of Eq.~\eqref{EulerI} are canceled, and we obtain 
\bea{\label{EulerT} &\partial_t (\rho v) + \partial_{\bar{z}} (\rho v v)+\partial_{z} (\rho v \bar{v})+\rho \partial_z(2p) \\
	&+4 \eta^2  \left \{ 4 \pi \partial_z \sigma^2+  \partial_{z}\left[ \sigma \partial_{\bar{z}}(\rho^{-1} \partial_z \sigma )\right] + \partial_{\bar{z}}\left[ \sigma \partial_{z} (\rho^{-1} \partial_z \sigma)\right] \right \}\nn\\
	&-4\eta i \partial_{\bar{z}}(\sigma \partial_z v)=0.\nn}
Introducing the complex momentum flux tensors
\bea{{\cal T}_{z\bar z}\equiv\rho v \bar v + 16 \eta^2 \pi \sigma^2+4 \eta^2\sigma \partial_{\bar z} (\rho^{-1}\partial_{z} \sigma),}
and
\bea{{\cal T} \equiv \rho v v +4 \eta^2 \sigma \partial_z (\rho^{-1}\partial_z \sigma)-4\eta i \sigma \partial_z v,} Eq.~\eqref{EulerT} becomes 
\bea{\partial_t(\rho v)+\partial_z {\cal T}_{z\bar z}+\partial_{\bar z} {\cal T}+\rho \partial_z(2p)=0,}
which is Eq.~\eqref{universalform} in the main manuscript. 

\subsection{Macroscopic Quantities}
In this section we show how to construct macroscopic quantities from the corresponding  microscopic quantities following a similar procedure in electromagnetism~\cite{Jackson}.  
The microscopic definition of charge density reads
\bea{\sigma(\mathbf{r})=\sum_i\sigma_i \delta(\mathbf{r}-\mathbf{r}_i).}
Considering a patch with characteristic length scale $\ell$ containing $M$ point vortices, the average charge density in the patch reads 
\bea{\label{average}\langle \sigma (\mathbf{r}) \rangle&=\ell^{-2}\int_{\ell^2}  d^2 \mathbf{r}' \ \sigma (\mathbf{r}-\mathbf{r}') \nn\\
	&= \ell^{-2} \int_{\ell^2} d^2 \mathbf{r}' \ \sum_{i}\sigma_i \delta (\mathbf{r}-\mathbf{r}_i-\mathbf{r}') \nn\\
	&= \ell^{-2}  \sum^{M}_{i} \sigma_i,}
Fluctuations on the scale smaller than $\ell$ are averaged out in Eq.~\eqref{average}, giving a relatively smooth macroscopic quantity on the scale of the patch size $\ell \gg \xi$. In this coarse-grained picture, a patch is viewed as a point $\mathbf{r}$ and the macroscopic fields describe collections of vortices at the patch. 
Similarly, for the vortex density
\bea{\label{rhoaverage}\langle \rho (\mathbf{r}) \rangle&= \ell^{-2}  M.} 
The coarse-grained, macroscopic vortex number current reads 
\bea{\langle J_{\rm n}(\mathbf{r}) \rangle &= \ell^{-2} \int_{\ell^2} d^2 \mathbf{r}' \ \sum_{i} v_i \delta (\mathbf{r}-\mathbf{r}_i-\mathbf{r}') \nn\\
	&=\ell^{-2} \sum_{i} v_i \int_{\ell^2} d^2 \mathbf{r}' \delta (\mathbf{r}-\mathbf{r}_i-\mathbf{r}') \nn\\
	&=\ell^{-2} \sum^M_{i=1} v_i.}
The relation $\langle J_{\rm n}  \rangle= \langle \rho \rangle v$ gives the vortex velocity field 
\bea{v=\langle \rho \rangle^{-1} \langle J_{\rm n}  \rangle=M^{-1}\sum^{M}_{i=1} v_i,}
which is used to calculate the average vortex velocities for different bin sizes, showing that the average vortex velocities $ M^{-1} \sum_i v_i$ approaches the continuous field $v$ with increasing the bin size (see~\fref{shearflowFig} in the main manuscript).

\subsection{Numerical Simulations of the Point vortex model and the Vortex Shear Flow}

In the simulations we use the dissipative PVM 
\bea{\label{dissipativePV}
	\frac{d\bar{z}_i}{dt}= v_i+i \eta^{\ast} \sigma_i v_i.} 
We consider a neutral point vortex system in a square box with side length $L$. 
Since the vortex shear flow described by Eq.~\eqref{shearflow} in the main manuscript is periodic in the $y$ direction and homogeneous in the $x$ direction, we impose periodic boundary conditions. In a doubly-periodic square box, 
the velocity of the $i$th vortex in the presence of the $j$th,  $\mathbf{v}_{i}^{(j)}$, reads~\cite{doublyperiodic}
\begin{equation}
\vv_{i}^{(j)} = 
\frac{ \kappa \sigma_j} {2 L}
\sum_{m=-\infty}^{\infty}
\left( \begin{array}{c}
\frac{-\sin(y_{ij}^\prime)}{\cosh(x_{ij}^\prime-2\pi m) - \cos(y_{ij}^\prime)}\\
\frac{\sin(x_{ij}^\prime)}{\cosh(y_{ij}^\prime-2\pi m) - \cos(x_{ij}^\prime)}
\end{array} \right),\label{eqn:periodic_v}
\end{equation}
where $\mathbf{r}_{ij}^\prime =(\rr_i -
\rr_j) (2\pi/L)$. 
To simulate the dynamics of large number of vortices in a doubly-periodic square box, we adopted the computational scheme developed in Ref.~\cite{Matt2017}. We measure length $L$, velocity $v$ and time $t$ in units of $\xi$, $\kappa/\xi$ and $\xi^2/\kappa$ respectively.

The annihilation of vortex--antivortex pairs is modelled by removing vortex dipoles with separation less than the healing length $\xi$. To take into account the energy dissipation due to the sound wave radiation of accelerating vortices, we increase the dissipation $\eta^{*}$ for the same-sign vortex pairs when their separation approaches the healing length $\xi$~\cite{Pismen}. In our simulations we use the same dissipation as in Ref.~\cite{Matt2017}:
\begin{equation}
\eta^{*}_i = \mathrm{max}\left( \exp \left[ \ln(\eta^{*}) \frac{r_{is}-r_1}{r_2-r_1} \right],\eta^{*} \right)\,,
\label{eqn:DissipationEnhancement}
\end{equation}
and we choose $r_2 = \xi$ and $r_1 = 0.1\xi$ without losing generality. When computing the evolution of vortex $i$, the background dissipation rate $\eta^{*}=10^{-4}$ in Eq.~\eqref{dissipativePV} is replaced with the local dissipation rate $\eta^{*}_i$. 
During our simulations we find that vortex number loss is weak and the highest loss ratio is around $3\%$.  

The initial vortex configuration is obtained by sampling vortex coordinates according to the distribution $\sigma=\sigma_0 \sin(2 \pi y/L)$, choosing $\sigma_0=\rho_0$. We sample positions of point vortices (antivortices) according to the distributions 
\bea{n_{\pm}=\frac{\rho_0}{2}\left[1\pm\sin\left(\frac{2\pi y}{L}\right)\right],
}
respectively, where $y\in [-L/2,L/2]$. We then make replicas of these vortices and distribute them uniformly along the $x$-axis. In our sampling the constant vortex density is anisotropic. Being anisotropic is irrelevant as long as the densities along the $x$-axis and the $y$-axis are defined at similar scales.    

We keep track of the coordinates of vortices initially located at $x=0$. From the trajectories of these vortices we can extract the vortex velocity field distribution.  There is no motion along the $y$ direction. We fit our results using the fitting function 
\bea{\label{fitting}v=v_0 \cos\left(\frac{2\pi}{L} y + \theta_0\right)+C.}
The fitting parameters $v_0$, $\theta_0$ and $C$ are shown in Table~\ref{table1}. The parameter $v_0$ barely changes if $\theta_0$ and $C$ are set to be zero for each $L$.

\renewcommand{\tabcolsep}{7pt}
\begin{table}[!t]
	\centering 
	\begin{tabular}{ccccc}
		\hline\hline
		$L\;[10^3]$ & $v_0 (\rm fitting)$ & $v_0 (\rm theory)$ & $\theta_0$ & $C$ \\
		\hline
		$1$ & $1.4265$ & $1.5155$ &   $0.0152$  &  $0.0820$ \\ 
		$2$ & $0.7811$ & $0.7577$ &  $0.0828$  &  $0.0043$\\
		$3$ & $0.5181$ & $ 0.5052$ &  $-0.0940$ & $-0.0563$  \\
		$4$ & $0.3778$ & $0.3789$ &  $-0.0296$ &  $0.0175$\\
		$5$ & $0.3163$ & $0.3031$ &  $-0.0094$  & $0.0047$\\
		$6$ & $0.2332$ & $ 0.2526$ &  $-0.0572$ &  $-0.0001$\\
		$7$ & $0.2143$ & $0.2165 $ &  $-0.0807$ & $-0.0050$ \\
		$8$ &  $0.1903$ & $0.1894$ &  $-0.1280$  & $0.0180$  \\
		$9$ & $0.1660$ & $  0.1684$ &  $0.2033$ & $-0.0106$ \\
		$10$ & $0.1567$ & $ 0.1515 $ & $0.1662$  & $-0.0029$  \\
		\hline\hline
		
	\end{tabular}
	\caption{Numerical fitting parameters for $N=9522$ and different $L$. For each $L$, the results are for a single run (time evolution for a single sample from the initial state distribution). For $L=\{1,2,3,4\}\times 10^3$ the fit is made at $t=2$, and for $L=\{5,6,7,8,9,10\}\times 10^3$ the fit is made at $t=50$.\label{table1}}
\end{table}

\begin{figure}[!t]
	\includegraphics[width=3in]{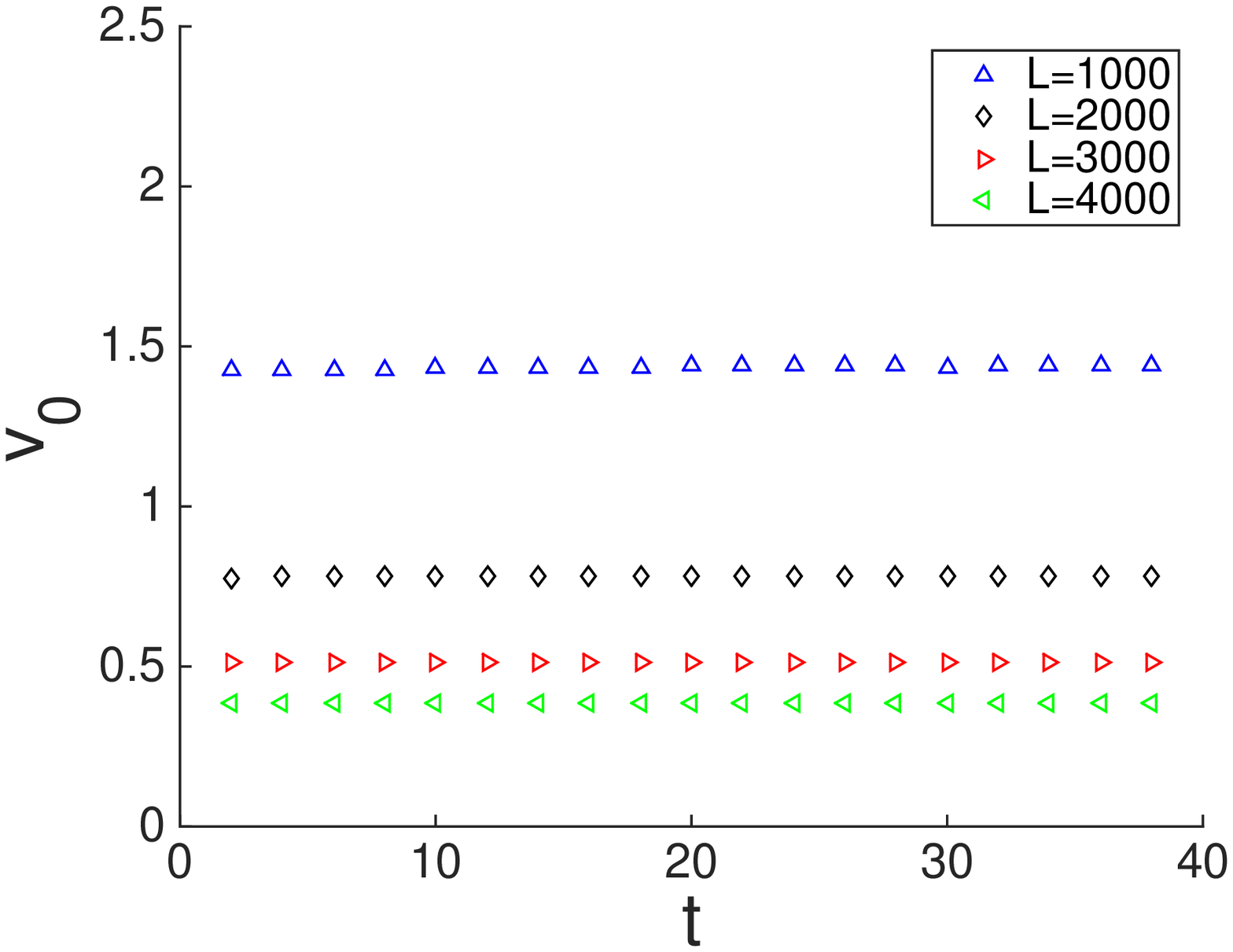}
	\includegraphics[width=3in]{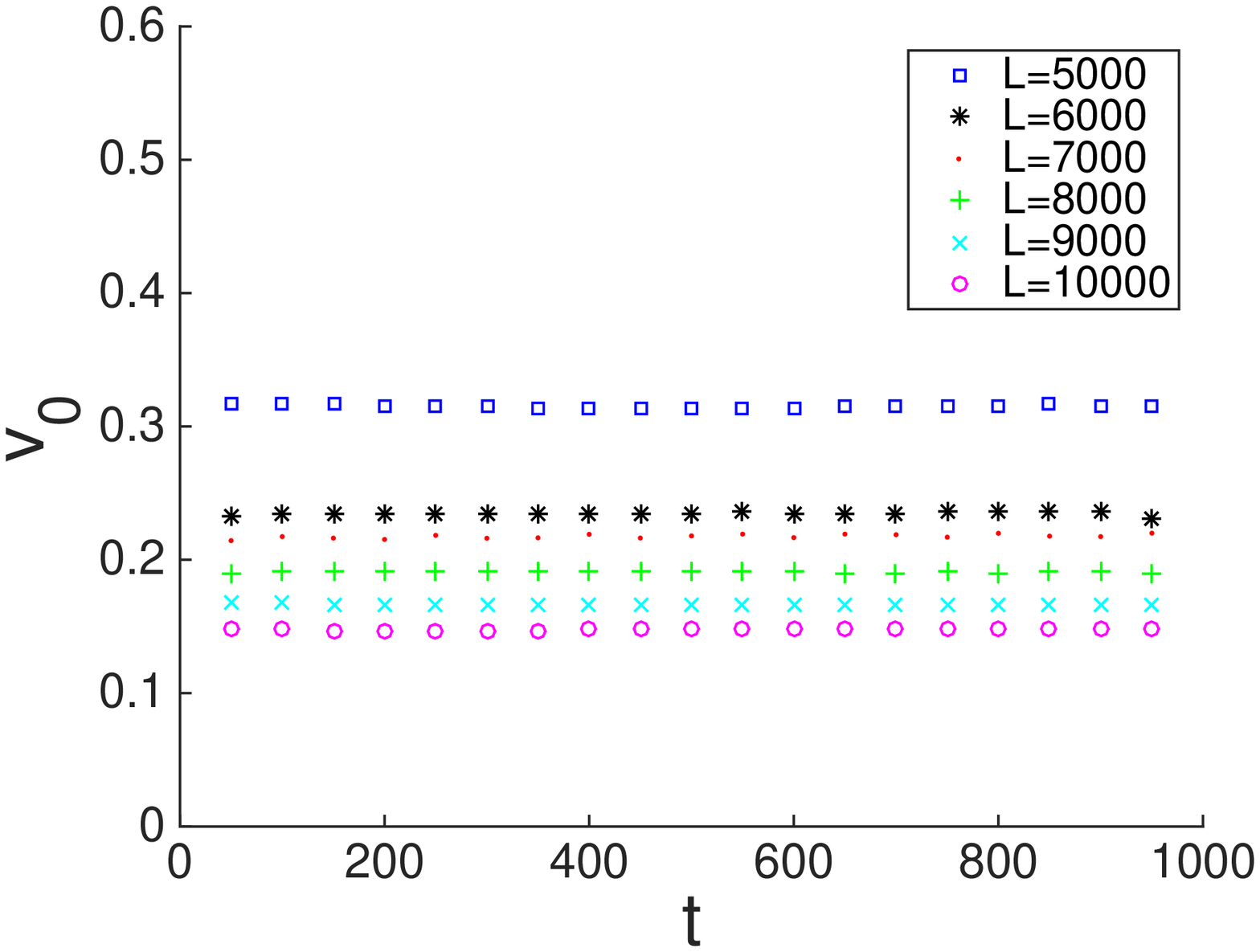}
	\caption{(Color online) The magnitude of the vortex velocity field $v_0$ fitted by Eq.~\eqref{fitting} from the numerical data at different times for $N=9522$ and different $L$. For each $L$, the data is collected every $2$ time units (top) and 
		$50$ time units (bottom) in a single run. 		
	} \label{velocity}
\end{figure}

\begin{figure}[!t]
	\includegraphics[width=3.3in]{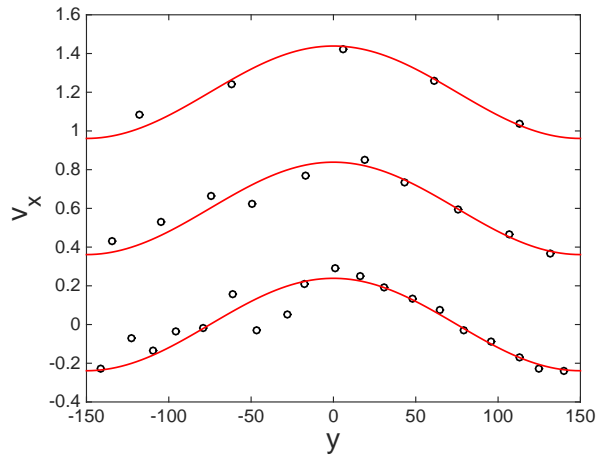}
	\caption{(Color online) A comparison between Eq. (25) in the main manuscript (red) and simulations (circles).  Same scenario as Fig.~1 (d) in the main manuscript, but for a much smaller system: $N=450$ and $L=300$. The vortex velocities are obtained as averages over $10$ runs. By averaging the velocities (bottom) over the increasing bin size $L/\{19,10,5\}$, the coarse-grained vortex velocities approach the analytical values (from bottom to top). For clarity the curves are each vertically shifted by 0.6. Comparing to Fig.1 (d) in the main manuscript, fluctuations of vortex velocities are stronger due to larger sampling fluctuations for a much smaller system.} \label{shearflowsmFig}
\end{figure}

For each $L$ the fitting parameter $v_0$ remains nearly constant over a long time interval (see Fig.~\ref{velocity}).  Fig.~\ref{shearflowsmFig} shows the coarse-grained vortex velocities for various bin sizes for a much smaller point vortex system and the results still show good agreement with the theory. 

\subsection{Vortex number Loss}

The proposed theory describes emergent properties of well-separated quantum votices at large scales and is valid whenever the PVM is applicable and the vortex number loss is low.
In various regimes of interest such conditions can be fulfilled and vortex losses are only a weak modification of the Hamiltonian vortex motion. In general, the collision rate depends on the vortex distribution and varies from case to case. For the vortex shear flow described in the main manuscript, our simulations of the dissipative PVM show that the vortex number loss is below $3 \% $. In the recent study on the enstrophy cascade based on the dissipative PVM, the annihilation rate is also shown to be negligible (less than $1 \% $)~\cite{Matt2017}. 

In a BEC vortex annihilation events involve physics outside the scope of the Hamiltonian PVM. The short range approach of vortices is associated with the formation of Jones-Roberts soliton~\cite{jones1982motions,jones1986motions}, a metastable density dip that occurs at the onset of the gradual process of complete annihilation into sound waves. The Jones-Roberts soliton forms at $d\approx 10\xi$~\cite{jones1986motions,SM7} marking the limit of validity of the PVM for BEC experiments.    
We estimate the annihilation rate for situations where vortex collision events are most frequent.   
In the absence of energy dissipation the vortex number loss  is dominated by vortex dipole annihilations~\cite{Barenghi2015,Gasenzer2016}, being significant when the system is in the homogeneous dipole gas regime, where collisions happen between vortex dipole pairs. In this regime the vortex collision cross-section $\sigma_{\rm cs}\sim \xi$. The root-mean-square vortex velocity $v_{\rm {rms}} \sim \kappa d^{-1} \sim \kappa\ (10\xi)^{-1}$, the mean free path ${\ell}_m \sim (\sigma_{\rm cs} \rho_{\rm d})^{-1}$, and the mean free time $\tau ={\ell}_m/ v_{\rm rms}\sim 10/(\kappa \rho_{\rm d})$, where $\rho_{\rm d}=\rho/2$ is the dipole density. Hence the average collision frequency for one vortex dipole is $Z=\tau^{-1}=\kappa \rho_{\rm d}/10$ and the average total collision frequency per unit area is $Z_t=Z \rho_{\rm d}$.
We assume that for every dipole-dipole collision there are two vortices being annihilated, then the vortex number loss during $\delta t$ is $\delta N(t)=N(t+\delta t)-N(t)=-2Z_t A \delta t$, where $A$ is the total area of the system.

Hence the decay rate equation reads 
\bea{\label{dipoledecay}\frac{d N}{dt}=-\Gamma_2 N^2=-\frac{1}{20}\kappa A^{-1}N^2,}
leading to $\rho (t) \sim t^{-1}$. Such vortex decay behavior has been observed in the recent experiment~\cite{Relaxation2014Shin} and is consistent with Refs.~\cite{Gasenzer2012pra,Barenghi2015}. Note that the pre-factor in Eq.~\eqref{dipoledecay} is over-estimated as some of the dipole-dipole scatterings may not evolve annihilation~\cite{Gasenzer2012pra}. Taking the parameters in the experiment on  $^{23}\rm Na$~\cite{Relaxation2014Shin}, where the atomic mass $m_{\rm a}=23 \times 1.7 \times 10^{-27} \rm {kg}$ and the radius of Thomas-Fermi density profile $R_{\rm TF}=7\times 10^{-5} m$,  we estimate the two-body decay rate $\Gamma_2=(1/20) \ \kappa A^{-1}\simeq 0.054 \ s^{-1}$. The result is comparable to the hottest experiment and is around $10$ times of the value for the coldest experiment in Ref.~\cite{Relaxation2014Shin}.
Here we used $A=\pi R^2_{\rm TF}$ and $\kappa=h/m_{\rm a}$ with $h\simeq 6.6\times 10^{-34} \rm {m^2 kg s^{-1}}$. For the same parameters the mean free time (lifetime) of a dipole is $\tau\sim 300{\rm ms}$, setting a time limit up to when the Hamiltonian PVM is a valid description.  

In above estimations, we assume that dipole-dipole scattering processes are dominant.  Vortex decay rates will be modified if three-body scattering processes (a vortex dipole scattering with a free vortex) are important~\cite{Gasenzer2016}, which could be the case when close to the Berezinskii-Kosterlitz-Thouless transition. It is worth mentioning that for a finite system, drift-out vortex number loss could be also relevant, while less important if energy dissipation is low ~\cite{Barenghi2015}. \\ 

\subsection{Experimental Accessibility}
Experiments have realized large atom-number condensates of up to $N_a \sim 10^8$ atoms~\cite{Streed2006,Straten}, with atomic
number densities $n_0 \sim 10^{14}$ cm$^{-3}$, using
$^{23}$Na~\cite{Streed2006}. The healing length $\xi
\approx 0.6 \mu\mathrm{m}$. If such a condensate can be realized in a uniform quasi-2D trap with side length $L$ and thickness $\sim 6.6\xi$~\cite{Relaxation2014Shin}, then 
this would allow that $L \sim 1000 \xi$ and the vortex number $N \sim 450 $. Since here
$u_{\mathrm{rms}}/c \lesssim 0.3$, such a system is within the incompressible regime.
Experiments in quasi-2D have
achieved $N\sim 60$ vortices for $L \sim 120 \xi$ in harmonically
trapped systems~\cite{Relaxation2014Shin}, and hard-wall confinement with $L \gtrsim 200 \xi$~\cite{Gauthier16}. With a lot of scope for larger condensates, and increasing experimental control, it is reasonable to expect conditions for which the hydrodynamic theory applies could be achieved in near-future BEC experiments.

\end{document}